\documentclass[onecolumn, 12pt, floatfix,aps,pra,superscriptaddress,nofootinbib]{revtex4}
\usepackage{graphicx}
\usepackage{amsmath}
\usepackage{amssymb}
\usepackage{bm}
\usepackage{url}
\usepackage{stmaryrd}
\usepackage{makecell}
\usepackage{setspace}
\usepackage[colorlinks=true, linkcolor=blue]{hyperref}
\newcommand{\nsys}{n_{\mathrm{sys}}}
\newcommand{\neig}{N_{\mathrm{eig}}}

\newcommand{\nfreq}{\#{\rm freq}}

\newcommand{\ucontrolled}{\mathcal{U}_c}

\newcommand{\Pfuncbold}{P_{\mathbf{k},\boldsymbol{\beta}}(\mathbf{m}|\boldsymbol{\phi},\mathbf{A})}

\newcommand{\Pfunc}{P_{\mathbf{k},\boldsymbol{\beta}}(\mathbf{m}|\phi)}
\newcommand{\Pfuncsr}{P_{k,\beta}(m|\phi)}

\newcommand{\<}{\langle}
\renewcommand{\>}{\rangle}

\newcommand{\II}{\mathbb{I}}
\newcommand{\Hh}{\mathcal{H}}

\begin{document}
\title[QPE of multiple eigenvalues for small-scale (noisy) experiments]{Quantum phase estimation of multiple eigenvalues for small-scale (noisy) experiments}
\author{Thomas E. O'Brien$^1$, Brian Tarasinski$^2$ and Barbara M. Terhal$^{2,3}$}
\address{$^1$ Instituut Lorentz, Universiteit Leiden, P.O. Box 9506, 2300 RA Leiden, The Netherlands \\
$^2$ QuTech, Delft University of Technology, P.O. Box 5046, 2600 GA Delft, The Netherlands \\
$^3$ JARA Institute for Quantum Information (PGI-11), Forschungszentrum Juelich, D-52425 Juelich, Germany}
\date{\today}
\begin{abstract}
Quantum phase estimation is the workhorse behind any quantum algorithm and a promising method for determining ground state energies of strongly correlated quantum systems.
Low-cost quantum phase estimation techniques make use of circuits which only use a single ancilla qubit, requiring classical post-processing to extract eigenvalue details of the system.
We investigate choices for phase estimation for a unitary matrix with low-depth noise-free or noisy circuits, varying both the phase estimation circuits themselves as well as the classical post-processing to determine the eigenvalue phases.
We work in the scenario when the input state is not an eigenstate of the unitary matrix.
We develop a new post-processing technique to extract eigenvalues from phase estimation data based on a classical time-series (or frequency) analysis and contrast this to an analysis via Bayesian methods.
We calculate the variance in estimating single eigenvalues via the time-series analysis analytically, finding that it scales to first order in the number of experiments performed, and to first or second order (depending on the experiment design) in the circuit depth. Numerical simulations confirm this scaling for both estimators.
We attempt to compensate for the noise with both classical post-processing techniques, finding good results in the presence of depolarizing noise, but smaller improvements in $9$-qubit circuit-level simulations of superconducting qubits aimed at resolving the electronic ground-state of a $H_4$-molecule.
\end{abstract}
\maketitle

\tableofcontents
\section{Introduction}
It is known that any problem efficiently solvable on a quantum computer can be formulated as eigenvalue sampling of a Hamiltonian or eigenvalue sampling of a sparse unitary matrix~\cite{WZ06}.
In this sense the algorithm of quantum phase estimation is the only quantum algorithm which can give rise to solving problems with an exponential quantum speed-up.
Despite it being such a central component of many quantum algorithms, very little work has been done so far to understand what quantum phase estimation offers in the current NISQ (Noisy Intermediate Scale Quantum) era of quantum computing~\cite{Pre18} where quantum devices are strongly coherence-limited.
Quantum phase estimation comes in many variants, but a large subclass of these algorithms (e.g.~the semi-classical version of textbook phase estimation~\cite{semiFT, Nie00}, Kitaev's phase estimation~\cite{Kit95}, Heisenberg-optimized versions~\cite{KLY:QPE}), are executed in an iterative sequential form using controlled-$U^k$ gates with a single ancilla qubit~\cite{KOS:QPE, Svo13} (see Fig.~\ref{fig:QPE_circuit}), or by direct measurement of the system register itself~\cite{KLY:QPE}.
Such circuits are of practical interest in the near term when every additional qubit requires a larger chip and brings in additional experimental complexity and incoherence.

Some of the current literature on quantum phase estimation works under limiting assumptions.
The first is that one does not start in an eigenstate of the Hamiltonian~\cite{Mcc17,Wie16}.
A second limitation is that one does not take into account the (high) temporal cost of running $U^k$~\cite{Svo13} for large $k$ when optimizing phase estimation.
The size and shallowness of the quantum phase estimation circuit is important since, in the absence of error correction or error mitigation, one expects entropy build-up during computation.
This means that circuits with large $k$ may not be of any practical interest.

The scenario where the input state is not an eigenstate of the unitary matrix used in phase estimation is the most interesting one from the perspective of applications, and we will consider it in this work.
Such an input state can be gradually projected onto an eigenstate by the phase estimation algorithm and the corresponding eigenvalue can be inferred.
However, for coherence-limited low-depth circuits one may not be able to evolve sufficiently long to project well onto one of the eigenstates.
This poses the question what one can still learn about eigenvalues using low-depth circuits.
An important point is that it is experimentally feasible to repeat many relatively shallow experiments (or perform them in parallel on different machines).
Hence we ask what the spectral-resolving power of such phase estimation circuits is, both in terms of the number of applications of the controlled-$U$ circuit in a single experiment, and the number of times the experiment is repeated.
Such repeated phase estimation experiments require classical post-processing of measurement outcomes, and we study two such algorithms for doing this.
One is our adaptation of the Bayesian estimator of~\cite{Wie16} to the multiple-eigenvalue scenario.
A second is a new estimator based on a treatment of the observed measurements as a time-series, and construction of the resultant time-shift operator. This latter method is very natural for phase estimation, as one interprets the goal of phase estimation as the reconstruction of frequencies present in the output of a temporal sound signal. In fact, the time-series analysis that we develop is directly related to what are called Prony-like methods in the signal-processing literature, see e.g. \cite{pt:prony}. The use of this classical method in quantum signal processing, including in quantum tomography \cite{OWE:RB}, seems to hold great promise.

One can interpret our results as presenting a new hybrid classical-quantum algorithm for quantum phase estimation. Namely, when the number of eigenstates in an input state is small, i.e. scaling polynomially with the number of qubits $\nsys$, the use of our classical post-processing method shows that there is no need to run a quantum algorithm which projects onto an eigenstate to learn the eigenvalues. We show that one can extract these eigenvalues efficiently by classically post-processing the data from experiments using a single-round quantum phase estimation circuits (see Section \ref{sec:QPE_def}) and classically handling ${\rm poly}(\nsys) \times {\rm poly}(\nsys)$ matrices.  This constitutes a saving in the required depth of the quantum circuits. 

The spectral-resolution power of quantum phase estimation can be defined by its scaling with parameters of the experiment and the studied system. We are able to derive analytic scaling laws for the problem of estimating single eigenvalues with the time-series estimator.
We find these to agree with the numerically-observed scaling of both studied estimators.
For the more general situation, with multiple eigenvalues and experimental error, we study the error in estimating the lowest eigenvalue numerically.
This is assisted by the low classical computation cost of both estimators.
We observe scaling laws for this error in terms of the overlap between the ground and starting state (i.e. the input state of the circuit), the gap between the ground and excited states, and the coherence length of the system.
In the presence of experimental noise we attempt to adjust our estimators to mitigate the induced estimation error.
For depolarizing-type noise we find such compensation easy to come by, whilst for a realistic circuit-level simulation we find smaller improvements using similar techniques.

Even though our paper focuses on quantum phase estimation where the phases corresponds to eigenvalues of a unitary matrix, our post-processing techniques may also be applicable to multi-parameter estimation problems in quantum optical settings. In these settings the focus is on determining an optical phase-shift \cite{xiang+:PE, daryanoosh+:PE, berni+:PE} through an interferometric set-up. There is experimental work on (silicon) quantum photonic processors \cite{Zhou+:QPE, QPE:exp-sil,santagati+:PE} on multiple-eigenvalue estimation for Hamiltonians which could also benefit from using the classical post-processing techniques that we develop in this paper. 

\section{Quantum phase estimation}\label{sec:QPE_def}
Quantum phase estimation (QPE) covers a family of quantum algorithms which measure a system register of $\nsys$ qubits in the eigenbasis of a unitary operator $U$~\cite{Kit95,Abr99}
\begin{equation}
U|\phi_j\>=e^{i\phi_j}|\phi_j\>,
\end{equation}
to estimate one or many phases $\phi_j$.
Quantum phase estimation algorithms assume access to a noisefree quantum circuit which implements $U$ on our system register conditioned on the state of an ancilla qubit.
Explicitly, we require the ability to implement
\begin{equation}
\ucontrolled = |0\> \<0|\otimes\II +|1\> \<1|\otimes U~\label{eqn:exact_circuit},
\end{equation}
where $|0\>$ and $|1\>$ are the computational basis states of the ancilla qubit, and $\II$ is the identity operator on the system register.

In many problems in condensed matter physics, materials science, or computational chemistry, the object of interest is the estimation of spectral properties or the lowest eigenvalue of a Hamiltonian $\Hh$.
The eigenvalue estimation problem for $\Hh$ can be mapped to phase estimation for a unitary $U_\tau=\exp(-i \tau \Hh)$ with a $\tau$ chosen such that the relevant part of the eigenvalue spectrum induces phases within $[-\pi,\pi)$.
Much work has been devoted to determining the most efficient implementation of the (controlled)-$\exp(- i \tau \Hh)$ operation, using exact or approximate methods~\cite{Whi09,Abr99,Ber15,Low17}.
Alternatively, one may simulate $\Hh$ via a quantum walk, mapping the problem to phase estimating the unitary $\exp(-i \arcsin(\Hh)/\lambda)$ for some $\lambda$, which may be implemented exactly~\cite{Ber17,Pou17,Ber12,Chi10}.
In this work we do not consider such variations, but rather focus on the error in estimating the eigenvalue phases of the unitary $U$ that is actually implemented on the quantum computer.
In particular, we focus on the problem of determining the value of a single phase $\phi_0$ to high precision (this phase could correspond, for example, to the ground state energy of some Hamiltonian $\Hh$).

Phase estimation requires the ability to prepare an input, or {\em starting state}
\begin{equation}
|\Psi\>=\sum_j a_j|\phi_j\>, A_j \equiv |a_j|^2 \label{eqn:psi},
\end{equation}
with good overlap with the ground state; $A_0 \gg 0$.
Note here that the spectrum of $U$ may have exact degeneracies (e.g.~those enforced by symmetry) which phase estimation does not distinguish; we count degenerate eigenvalues as a single $\phi_j$ throughout this work.
The ability to start quantum phase estimation in a state which already has good overlap with the ground state is a nontrivial requirement for the applicability of the quantum phase estimation algorithm.
On the other hand, it is a well-known necessity given the QMA-completeness \cite{Kit02} of the lowest eigenvalue problem~\footnote{QMA stands for Quantum Merlin Arthur, which is a complexity class which contains decision problems which are easy to verify on a quantum computer, though not necessarily easy to solve. This class is the natural quantum counterpart to the complexity class NP of problems that may be verified easily on a classical computer. A QMA-complete problem is one of the `hardest possible' such problems (in analogy with NP-complete problems); the ability to solve these problems in polynomial time would allow polynomial-time solving of any other problem in QMA.}.
For many quantum chemistry and materials science problems it is known or expected that the Hartree-Fock state has good overlap with the ground state, although rigorous results beyond perturbation theory are far and few between (see e.g.~\cite{BGKT}).
Beyond this, either adiabatic evolution~\cite{Whi09,Rei17} or variational quantum eigensolvers~\cite{Per14} can provide an approximate starting state to improve on via phase estimation.

Phase estimation is not limited to simply learning the value of $\phi_0$; it may obtain information about all phases $\phi_j$ as long as $A_j>0$.
However, the resources required to estimate $\phi_j$ are bounded below by $1/A_j$.
To see this, note that the controlled-unitary $\ucontrolled$ does not mix eigenstates, and so there is no difference (in the absence of error) between starting with $|\Psi\>$ and the mixed state
\begin{equation}
\rho_{\Psi}=\sum_jA_j|\phi_j\>\<\phi_j|.\label{eq:rhopsi}
\end{equation}
The latter is then equivalent to preparing the pure state $|\phi_j\>$ with probability $A_j$, so if $N$ preparations of $|\phi_j\>$ are required to estimate $\phi_j$ to an error $\epsilon$, the same error margin requires at least $N/A_j$ preparations of the state $|\Psi\>$.
As the number of eigenstates $\neig$ with non-zero contribution to $|\Psi\>$ generally scales exponentially with the system size $\nsys$, estimating more than the first few $\phi_j$ (ordered by the magnitude $A_j$) will be unfeasible.

\begin{figure}
\includegraphics[width=\columnwidth]{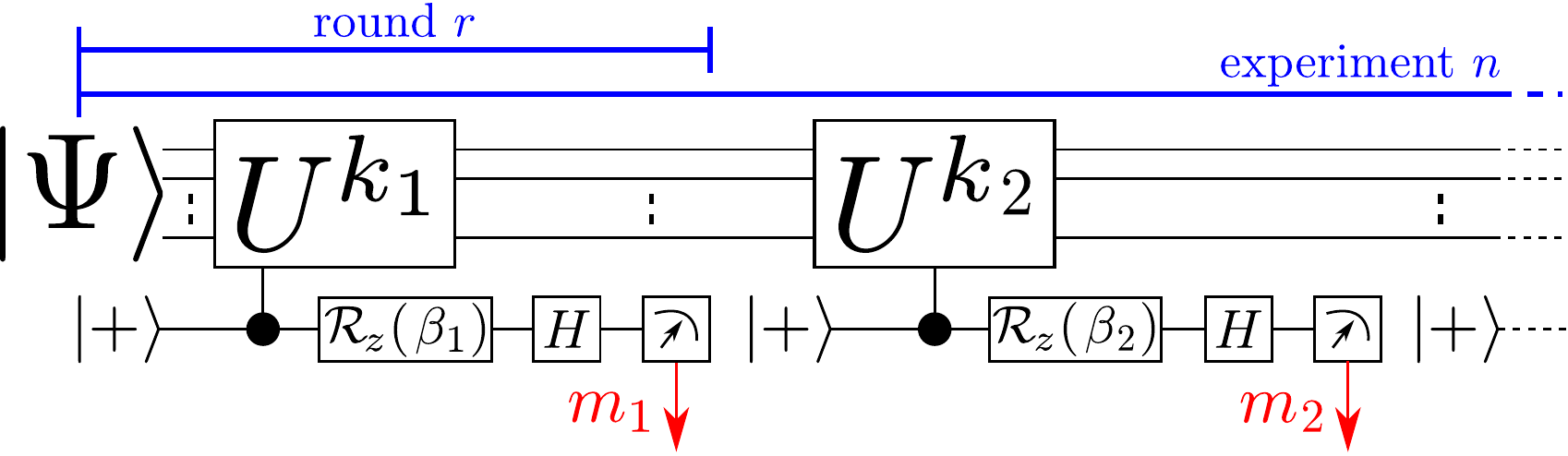}
\caption{\label{fig:QPE_circuit}Circuit for the QPE experiments described in this work. The state $|\Psi\>$ is defined in Eq.~(\ref{eqn:psi}). The probability for the ancilla qubit to return the vector $\mathbf{m}$ of results in the absence of error is given by Eq.~(\ref{eqn:QPE_probability_noerror}). The single-qubit rotation equals ${\cal R}_z(\beta)=\exp(-i \beta Z/2)$ while $H$ is the Hadamard gate.}
\end{figure}

Low-cost (in terms of number of qubits) quantum phase estimation may be performed by entangling the system register with a single ancilla qubit~\cite{Kit95,Kit02,Svo13,Wie16}.
In Fig.~\ref{fig:QPE_circuit}, we give the general form of the quantum circuit to be used throughout this paper.
An experiment, labeled by a number $n=1,\ldots,N$, can be split into one or multiple rounds $r=1,\ldots, R_n$, following the preparation of the starting state $|\Psi\>$.
In each round a single ancilla qubit prepared in the $|+\>=\frac{1}{\sqrt{2}}(|0\>+|1\>)$ state controls $\ucontrolled^{k_r}$ where the integer $k_r$ can vary per round.
The ancilla qubit is then rotated by ${\cal R}_z(\beta_r)=\exp(-i \beta_r Z/2)$ (with the phase $\beta_r$ possibly depending on other rounds in the same experiment) and read out in the $X$-basis, returning a measurement outcome $m_r \in \{0,1\}$.
We denote the chosen strings of integers and phases of a single multi-round experiment by $\bf{k}$ and $\boldsymbol{\beta}$ respectively.
We denote the number of controlled-$U$ iterations per experiment as $K=\sum_{r=1}^{R_n} k_r$.
We denote the total number of controlled-$U$ iterations over all experiments as
\begin{equation}
K_{\rm tot}=\sum_{n=1}^{N} \sum_{r=1}^{R_n} k_r.\label{eq:def_ktot}
\end{equation}
As the system register is held in memory during the entire time of the experiment, the choice of $K$ is dictated by the coherence time of the underlying quantum hardware.
Hence, we introduce a dimensionless coherence length 
\begin{align}
K_{\rm err}= \frac{T_{\rm err}}{\nsys T_U}.
\label{eq:def-coh}
\end{align}
Here $T_U$ is the time required to implement a single application of controlled-$U$ in Eq.~(\ref{eq:contU}), and $T_{\rm err}$ is the time-to-error of a single qubit, so that $T_{\rm err}/\nsys$ is the time-to-failure of $\nsys$ qubits.
The idea is that $K_{\rm err}$ bounds the maximal number of applications of $U$ in an experiment, namely $K \leq K_{\rm err}$.

A new experiment starts with the same starting state $|\Psi\>$.
Values of $k_r$ and $\beta_r$ may be chosen independently for separate experiments $n$, i.e.~we drop the label $n$ for convenience. We further drop the subscript $r$ from single-round experiments (with $R=1$).

In the absence of error, one may calculate the action of the QPE circuit on the starting state (defined in Eq.~(\ref{eqn:psi})). Working in the eigenbasis of $U$ on the system register, and the computational basis on the ancilla qubit, we calculate the state following the controlled-rotation $\ucontrolled^{k_1}$, and the rotation ${\cal R}_z(\beta_1)$ on the ancilla qubit to be
\begin{equation}
\frac{1}{\sqrt{2}}\sum_ja_j\left(|0\>+e^{i(k_1\phi_j+\beta_1)}|1\>\right)|\phi_j\>.
\label{eq:contU}
\end{equation}
The probability to measure the ancilla qubit in the $X$-basis as $m_1\in\{0,1\}$ is then
\begin{equation}
\sum_j A_j \cos^2\left(\frac{k_1\phi_j}{2}+\frac{\beta_1-m_1 \pi}{2}\right),
\end{equation}
and the unnormalized post-selected state of the system register is
\begin{equation}
\sum_ja_je^{\frac{i}{2}(k_1\phi_j+\beta_1)}\cos\left(\frac{k_1\phi_j}{2}+\frac{\beta_1-m_1 \pi}{2}\right)|\phi_j\>.
\end{equation}
The above procedure may then be repeated for $r$ rounds to obtain the probability of a string $\mathbf{m}$ of measurement outcomes of one experiment as
\begin{eqnarray}
\Pfuncbold = \sum_j A_j \prod_{r=1}^{R} \cos^2\left(\frac{k_r\phi_j}{2}+\frac{\beta_r-m_r\pi}{2}\right). \nonumber \\
\label{eqn:QPE_probability_noerror}
\end{eqnarray}
Here, $\boldsymbol{\phi}$ is the vector of phases $\phi_j$ and $\boldsymbol{A}$ the vector of probabilities for different eigenstates.
We note that $\Pfuncbold$ is independent of the order in which the rounds occur in the experiment.
Furthermore, when $\neig=1$, $\Pfunc=\Pfuncbold$ is equal to the product of the single-round probabilities $P_{k_r,\beta_r}(m_r|\phi)$, as there is no difference between a multi-round experiment and the same rounds repeated across individual experiments.

One can make a direct connection with parameter estimation work by considering the single-round experiment scenario in Fig.~\ref{fig:QPE_circuit}. The Hadamard gate putting the ancilla qubit in $|+\rangle$ and measuring the qubit in the $X$-basis are, in the optical setting, realized by beam-splitters, so that only the path denoted by the state $|1\rangle$ will pick up an unknown phase-shift. When the induced phase-shift is not unique but depends, say, on the state of another quantum system, we may like to estimate all such possible phases corresponding to our scenario of wishing to estimate multiple eigenvalues. 
Another physical example is a dispersively coupled qubit-cavity mode system where the cavity mode occupation number will determine the phase accumulation of the coupled qubit \cite{QED:num}. 

\section{Classical data analysis}\label{sec:data_analysis}
Two challenges are present in determining $\phi_0$ from QPE experiments. First, we only ever have inexact sampling knowledge of $\Pfuncbold$.
That is, repeated experiments at fixed $\mathbf{k},\boldsymbol{\beta}$ do not directly determine $\Pfuncbold$, but rather sample from the multinomial distribution $\Pfuncbold$.
From the measurement outcomes we can try to estimate $\Pfuncbold$ (and from this $\phi_0$) as a hidden variable.
Secondly, when $\neig>1$ determining $\phi_0$ from $\Pfuncbold$ poses a non-trivial problem.

Let us first consider the case $\neig=1$.
Let us assume that we do single-round experiments with a fixed $k$ for each experiment.
Naturally, taking $k=1$ would give rise to the lowest-depth experiments.
If we start these experiments with $k=1$ in the eigenstate $|\phi_0\rangle$, then one can easily prove that taking $\beta=0$ or $\beta=\frac{\pi}{2}$ for half of the experiments, suffices to estimate $\phi_0$ with variance scaling as $\sim 1/N=1/K_{\rm tot}$.
This result can be derived using standard Chernoff bounds, see e.g. \cite{higg:PE,Ter16}, and represent standard sampling or shot noise behavior.
When $\neig=1$, $N$ $K$-round experiments each with $k=1$ are indistinguishable from $N\times K$ single-round experiments with $k=1$.
This implies that the same scaling holds for such multi-round experiments, i.e. the variance scales as $1/(NK)=1/K_{\rm tot}$.

Once the phase $\phi_0$ is known to sufficient accuracy, performing QPE experiments with $k > 1$ is instrumental in resolving $\phi_0$ in more detail, since the probability of a single-round outcome depends on $k \phi_0$~\cite{KLY:QPE}.
Once one knows with sufficient certainty that $\phi_0\in[(2m-1)\pi/k,(2m+1)\pi/k)$ (for integer $m$), one can achieve variance scaling as $O(\frac{1}{k^2N})$ (conforming to so-called local estimation Cramer-Rao bounds suggested in~\cite{Wie16,Wie15}).
A method achieving Heisenberg scaling, where the variance scales as $\sim 1/K_{\rm tot}^2$ (see Eq.~(\ref{eq:def_ktot})), was analyzed in~\cite{KLY:QPE, higg:PE}.
This QPE method can also be compared with the information-theoretic optimal maximum-likelihood phase estimation method of~\cite{Svo13} where $N\sim \log K$ experiments are performed, each choosing a random $k \in \{1,\ldots, K\}$ to resolve $\phi_0$ with error $\sim 1/K$.
The upshot of these previous results is that, while the variance scaling in terms of the total number of unitaries goes like $1/K_{\rm tot}$ when using $k=1$, clever usage of $k >1$ data can lead to $1/K_{\rm tot}^2$ scaling.
However, as $K$ is limited by $K_{\rm err}$ in near-term experiments, this optimal Heisenberg scaling may not be accessible. 

When $\neig>1$, the above challenge is complicated by the need to resolve the phase $\phi_0$ from the other $\phi_j$. This is analogous to the problem of resolving a single note from a chord.
Repeated single-round experiments at fixed $k$ and varying $\beta$ can only give information about the value of the function:
\begin{equation}
g(k)=\sum_jA_je^{ik\phi_j},
\label{eq:defg}
\end{equation}
at this fixed $k$, since
\begin{align}
\Pfuncsr=&\frac{1}{2}+\frac{1}{2}\cos(\beta+m\pi)\mathrm{Re}[g(k)]\nonumber\\&-\frac{1}{2}\sin(\beta+m\pi)\mathrm{Im}[g(k)].
\end{align}
This implies that information from single-round experiments at fixed $k$ is insufficient to resolve $\phi_0$ when $\neig>1$, as $g(k)$ is then not an invertible function of $\phi_0$ (Try to recover a frequency from a sound signal at a single point in time!).
In general, for multi-round experiments using a maximum of $K$ total applications of $\ucontrolled$, we may only ever recover $g(k)$ for $k\leq K$.
This can be seen from expanding $\Pfuncbold$ as a sum of $\sum_jA_j\cos^m(\phi_j)\sin^n(\phi_j)$ terms with $m+n\leq K$, which are in turn linear combinations of $g(k)$ for $k\leq K$.
As we will show explicitly in the next Section \ref{sec:time} this allows us to recover up to $K$ $\phi_j$.
However, when $\neig>K$, these arguments imply that we cannot recover any phases exactly.
In this case, the accuracy to which we can estimate our target $\phi_0$ is determined by the magnitude of the amplitude $A_0$ in the inital state $|\Psi\>$ as well as the gap towards the other eigenvalues.
For example, in the limit $A_0\rightarrow 1$, an unbiased estimation of $\phi_0$ using data from $k=1$ would be
\begin{equation}
\mathrm{Arg}[g(1)]=\mathrm{Im}[\ln(\sum_jA_je^{i\phi_j})],
\end{equation}
and the error in such estimation is
\begin{align*}
|\mathrm{Arg}[g(1)]- \phi_0|&=|\frac{1}{A_0}\sum_{j=1}^{\neig-1}A_j\sin(\phi_j-\phi_0)+O(A_0^{-2})| \\
&\leq \frac{1-A_0}{A_0},
\end{align*}
with our bound being independent of $\neig$.
We are unable to extend this analysis beyond the $k=1$ scenario, and instead we study the scaling in this estimation numerically in Sec.~\ref{sec:num}.
In the remainder of this section, we present two estimators for multi-round QPE. The first is an estimator based on a time-series analysis of the function $g(k)$ using Prony-like~\cite{pt:prony} methods that has a low computation overhead.
The second is a Bayesian estimator similar to that of~\cite{Wie16}, but adapted for multiple eigenphases $\phi_j$. 

\subsection{Time-series analysis}\label{sec:time}
Let us assume that the function $g(k)$ in Eq.~(\ref{eq:defg}) is a well-estimated function at all points $0\leq k\leq K$, since the number of experiments $N$ is sufficiently large.
We may extend this function to all points $-K\leq k\leq K$ using the identity $g(-k)=g^*(k)$ to obtain a longer signal~\footnote{Extending $g(k)$ from $0\leq k\leq K$ to $-K\leq k\leq K$ is not required to perform a time-series analysis, however numerically we observe that this obtains up to order of magnitude improvement in estimating $\phi_0$.}.
We wish to determine the dominant frequencies $\phi_j$ in the signal $g(k)$ as a function of `time' $k$.
This can be done by constructing and diagonalizing a time-shift matrix $\frak{T}$ whose eigenvalues are the relevant frequencies in the signal, as follows.

We first demonstrate the existence of the time-shift matrix $\frak{T}$ in the presence of $\neig < K$ separate frequencies.
Since we may not know $\neig$, let us first estimate it as $l$.
We then define the vectors $\mathbf{g}(k)=(g(k),g(k+1),\ldots g(k+l))^T$, $k=-K,\ldots, K$.
These vectors can be decomposed in terms of single-frequency vectors $\mathbf{b}_j=(1,e^{i\phi_j},\ldots,e^{i l \phi_j})^T$
\begin{equation}
\mathbf{g}(k)=\sum_jA_j e^{i k\phi_j} \mathbf{b}_j.~\label{eq:gk_decomp}
\end{equation}
We can make a $l\times\neig$ matrix $B$ with the components $\mathbf{b}_j$ as columns
\begin{equation}
B_{k,j}=e^{ik\phi_j}.\label{eq:B_def}
\end{equation}
When $\neig \leq l$, the columns of $B$ are typically linearly independent~\footnote{Counterexamples may exist, but are hard to construct and have not occurred in any numerics.}, hence the non-square matrix $B$ is invertible and has a (left)-pseudoinverse $B^{-1}$ such that $B^{-1}B=\mathbf{1}$.
Note however, when $\neig > l$ the columns of $B$ are linearly-dependent, so $B$ cannot be inverted.
If $B$ is invertible, we can construct the shift matrix $\frak{T}=BDB^{-1}$ with $D_{i,j}=\delta_{i,j}e^{i\phi_j}$.
By construction, $\frak{T}\mathbf{b}_j=e^{i\phi_j}\mathbf{b}_j$ (as $\frak{T}B=BD$), and thus
\begin{align}
\frak{T}\mathbf{g}(k)&=\sum_jA_je^{ik\phi_j}\frak{T}\mathbf{b}_j\nonumber\\&=\sum_jA_je^{i(k+1)\phi_j}=\mathbf{g}(k+1).
\end{align}
This implies that $\frak{T}$ acts as the time-shift operator mapping ${\bf g}(k)$ to ${\bf g}(k+1)$. As the eigenvalues of $\frak{T}$ are precisely the required phases $e^{i\phi_j}$ in case $\neig \leq l$, constructing and diagonalizing $\frak{T}$ will obtain our desired phases including $\phi_0$.
When $\neig > l$, the eigen-equation for $\frak{T}$ cannot have the solution $\mathbf{b}_j$ since these are not linearly independent.

The above proof of existence does not give a method of constructing the time-shift operator $\frak{T}$, as we do not have access to the matrices $B$ or $D$.
To construct $\frak{T}$ from the data that we do have access to, we construct the $l \times (2K+1-l)$ Hankel matrices $G^{(0)}$, $G^{(1)}$ by
\begin{equation}
G^{(a)}_{i,j}=g(i+j+a-K),
\end{equation}
indexing $0 \leq i \leq l-1$, $0 \leq j \leq 2K-l$.
The $k$-th column of $G^{(a)}$ is the vector $\mathbf{g}(k+a-K)$, and so $\frak{T}G^{(0)}=G^{(1)}$.
We can thus attempt to find $\frak{T}$ as a solution of the (least-squares) problem of minimizing $||\frak{T}G^{(0)}-G^{(1)}||$.
The rank of the obtained $\tilde{\frak{T}}$ is bounded by the rank of $G^{(0)}$.
We have that $\mathrm{rank}(G^{(0)})$ is at most $\neig$ since it is a sum over rank-1 matrices.
At the same time $\mathrm{rank}(G^{(0)})\leq \min(l,2K+1-l)$.
This implies that we require both $l\geq\neig$ and $2K+1-l\geq\neig$ to obtain a shift matrix $\frak{T}$ with $\neig$ eigenvalues.
This is only possible when $K\geq\neig$, giving an upper bound for the number of frequencies obtainable.
When $G^{(0)}$ is not full rank (because $\neig<l$), this problem may have multiple zeros $\tilde{\frak{T}}$.
However, when $\neig<l$ each of these must satisfy $\tilde{\frak{T}}\mathbf{g}(k)=\mathbf{g}(k+1)$ for $-K<k< K-l$.

Then, as long as $\mathrm{rank}(G^{(0)})\geq\neig$, Eq.~(\ref{eq:gk_decomp}) is invertible by an operator $C$
\begin{equation}
\sum_{k}C_{i,k}A_je^{ik\phi_j}=\delta_{i,j}\rightarrow\mathbf{b}_j=\sum_{k}C_{j,k}\mathbf{g}(k).
\end{equation}
It follows that
\begin{equation}
\sum_k C_{j,k}\mathbf{g}(k+1)=\sum_{k,l} C_{j,k}A_le^{ik\phi_l}(e^{i\phi_l}\mathbf{b}_l)=e^{i\phi_j}\mathbf{b}_j,
\end{equation}
and then
\begin{equation}
\tilde{\frak{T}}\mathbf{b}_j=\sum_{k}C_{k,j}\tilde{\frak{T}}\mathbf{g}(k)=\sum_{k}C_{k,j}\mathbf{g}(k+1)=e^{i\phi_j}\mathbf{b}_j,
\end{equation}
so every $\tilde{\frak{T}}$ obtained in this way must have eigenvalues $e^{i\phi_j}$.

The above analysis is completely independent of the coefficients $A_j$.
However, once the eigenvalues $\phi_j$ are known, the matrix $B$ (eq.~\ref{eq:B_def}) may be constructed, and the $A_j$ may be recovered by a subsequent least-squares minimization of
\begin{equation}
||B\mathbf{A}-\mathbf{g}(0)||.~\label{eq:amp_min}
\end{equation}
This allows us to identify spurious eigenvalues if $l>\neig$ (as these will have a corresponding zero amplitude).
Numerically, we find no disadvantage to then choosing the largest $l$ permitted by our data, namely $l=K$.

\begin{figure}
\includegraphics[width=\columnwidth]{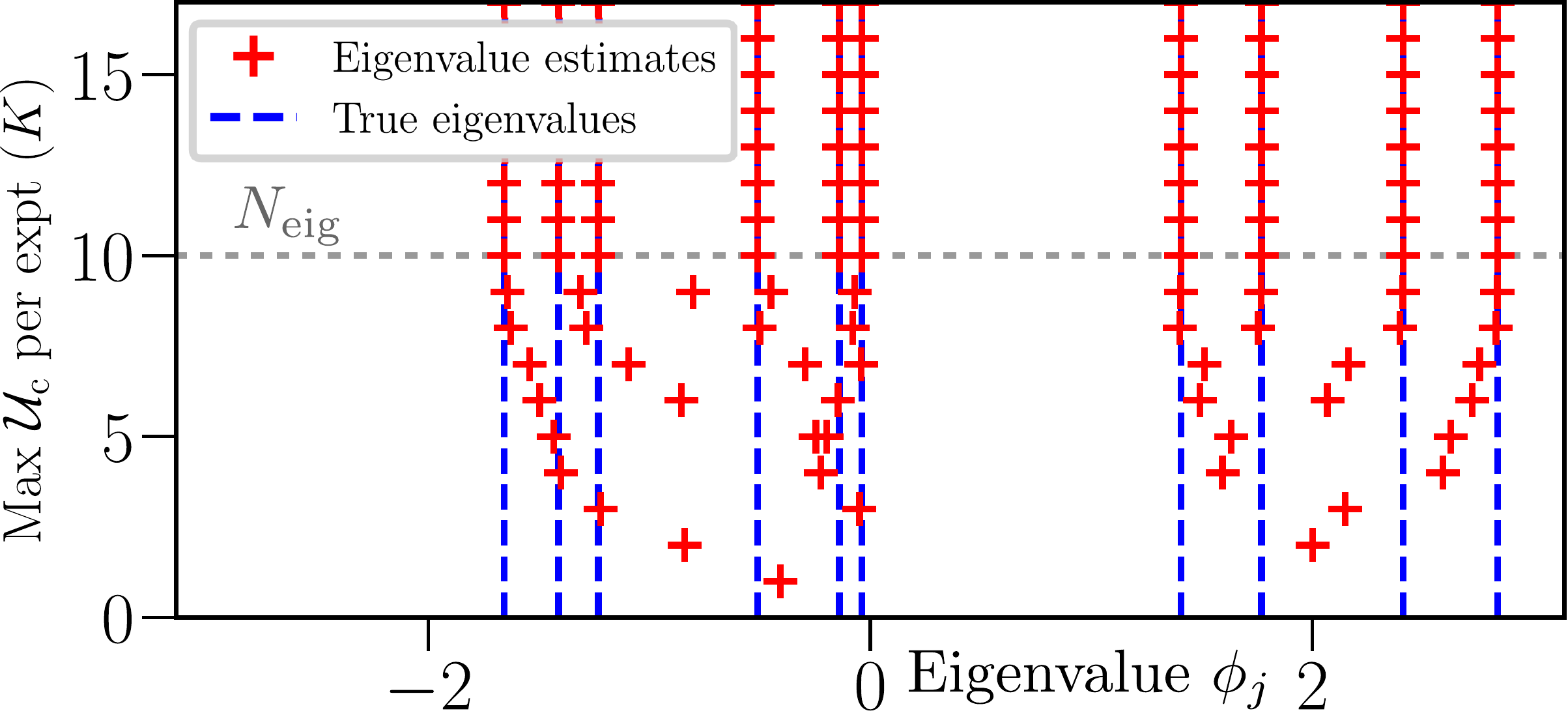}
\caption{\label{fig:TSE_performance}Convergence of the time-series estimator in the estimation of $\neig=10$ eigenvalues (chosen at random with equally sized amplitudes $A_j=1/10$) when the exact function $g(k)$ is known at points $0,\ldots,K$. The estimator constructs and calculates the eigenvalues of $K \times K$ matrix which are shown as the red plusses in the Figure. When $K\geq\neig$ (gray dashed line), the frequencies are attained to within machine precision. When $K < \neig$, it is clear from the Figure that the found eigenvalues provide some form of binning approximation of the spectrum.}
\end{figure}

Assuming a sufficient number of repetitions $N$ these arguments imply that this strategy requires that $K\geq\neig$ to determine all eigenvalues accurately.
However, when $K < \neig$ there still exists a least-squares solution $\tilde{\frak{T}}$ that minimizes $||\tilde{\frak{T}}G^{(0)}-G^{(1)}||$.
When $A_0\gg 0$, we expect that $\tilde{\frak{T}}$ should have eigenvalues $e^{i\tilde{\phi}_0}\approx e^{i\phi_0}$ that we can take as the estimator for $\phi_0$; the same is true for any other $\phi_j$ with sufficiently large $A_j$.
In Fig.~\ref{fig:TSE_performance} we show an example of convergence of this estimation for multiple eigenvalues $\phi_j$
 as $K\rightarrow\neig$ in the case where $g(k)$ is known precisely (i.e. in the absence of sampling noise).
The error $|\tilde{\phi}_0-\phi_0|$ when $K < \neig$ depends on the eigenvalue gap above $\phi_0$, as well as the relative weights $A_j$, as we will see in Section~\ref{sec:2ev}.

In \ref{app:calc_variance_full} we derive what variance can be obtained with this time-series method in the case $\l=\neig=1$, using single-round circuits with $k=1$ up to $K$.
Our analysis leads to the following scaling in $N$ and $K$:
\begin{equation}
\mathrm{Var}(\phi)\propto\frac{1}{K^2N}.\label{eq:time_series_var}
\end{equation}
We will compare these results to numerical simulations in Sec.~\ref{sec:1ev}.

\subsubsection{Estimating $g(k)$}~\label{sec:time_mr}

The function $g(k)$ cannot be estimated directly from experiments, but may instead be created as a linear combination of $\Pfuncbold$ for different values of $\mathbf{k}$ and $\mathbf{\beta}$.
For single-round experiments, this combination is simple to construct:
\begin{align}
g(k)=&P_{k,0}(0|\boldsymbol{\phi},\mathbf{A})-P_{k,0}(1|\boldsymbol{\phi},\mathbf{A})\nonumber\\&-iP_{k,\frac{\pi}{2}}(0|\boldsymbol{\phi},\mathbf{A})+iP_{k,\frac{\pi}{2}}(1|\boldsymbol{\phi},\mathbf{A}).\label{eq:gk_from_Pfunc}
\end{align}

For multi-round experiments, the combination is more complicated.
In general, $\Pfuncbold$ is a linear combination of real and imaginary parts of $g(l)$ with $l<K=\sum_rk_r$.
This combination may be constructed by writing $\cos^2(k\phi_j/2+\beta/2)$ and $\sin^2(k\phi_j/2+\beta/2)$ in terms of exponentials, and expanding.
However, inverting this linear equation is a difficult task and subject to numerical imprecision.
For some fixed choices of experiments, it is possible to provide an explicit expansion.
Here we focus on $K$-round $k=1$ experiments with $K/2$ $\beta=0$ and $K/2$ $\beta=\frac{\pi}{2}$ final rotations during each experiment (choosing $K$ even).
The formula for $\Pfuncbold$ is independent of the order in which these rounds occur.
Let us write $\mathbb{P}(\frak{m},\frak{n}|\boldsymbol{\phi},\mathbf{A})$ as the probability of seeing both 
$\frak{m} \in\{0,\ldots,K/2\}$ outcomes with $m_r=1$ in the $K/2$ rounds with $\beta_r=0$ and $\frak{n} \in\{0,\ldots,K/2\}$ outcomes with $n_r=1$ in the $K/2$ rounds with $\beta_r=\pi/2$.
In other words, $\frak{m}$, $\frak{n}$ are the Hamming weights of the measurement vectors split into the two types of rounds described above.
Then, one can prove that, for $0 \leq k \leq K/2$:
\begin{align}
g(k)=\sum_{m=0}^{K/2}\sum_{n=0}^{K/2} \chi_k (\frak{m},\frak{n})\mathbb{P}(\frak{m},\frak{n}|\boldsymbol{\phi},\mathbf{A})
\label{eq:exp-express}
\end{align}
where
\begin{align}
\chi_k(\frak{m},\frak{n})&=\sum_{l=0}^k (-i)^{k-l} {k \choose l} \nonumber \\
&\times\left[\sum_{p_1=0}^{\lfloor l/2 \rfloor} \frac{{\frak{m} \choose 2p_1} {K/2-\frak{m} \choose l-2p_1} }{{K/2 \choose l}}-1\right] \nonumber \\
&\times\left[\sum_{p_2=0}^{\lfloor (k-l)/2 \rfloor} \frac{{\frak{n} \choose 2p_2} {K/2-\frak{n} \choose k-l-2p_2}}{{K/2 \choose k-l}}-1 \right].
\label{eq:chi}
\end{align}
The proof of this equality can be found in \ref{app:equal}.

Calculating $g(k)$ from multi-round ($k=1$) experiments contains an additional cost: combinatorial factors in Eq.~(\ref{eq:exp-express}) relate the variance in $g(k)$ to the variance in $\mathbb{P}(\frak{m},\frak{n}|\boldsymbol{\phi},\mathbf{A})$ but the combinatorial pre-factor ${k \choose l}$ can increase exponentially in $k$.
This can be accounted for by replacing the least squares fit used above with a weighted least squares fit, so that one effectively relies less on the correctness of $g(k)$ for large $k$.
To do this, we construct the matrix $\frak{T}$ row-wise from the rows $\mathbf{g}^{(1)}_i$ of $G^{(1)}$.
That is, for the $i$th row $\mathbf{\frak{t}}_i$ we minimize
\begin{equation}
||\mathbf{\frak{t}}_i G^{(0)}-\mathbf{g}^{(1)}_i||.
\end{equation}
This equation may be weighted by multiplying $G^{(0)}$ and $g^{(1)}_i$ by the weight matrix
\begin{equation}
w^{(i)}_{j,k}=\delta_{j,k}\frac{1}{\sigma_{G^{(1)}_{i,j}}},\label{eq:weight_function}
\end{equation}
where $\sigma_{G^{(1)}_{i,j}}$ is the standard deviation in our estimate of $G^{(1)}_{i,j}$.
Note that the method of weighted least-squares is only designed to account for error in the independent variable of a least squares fit, in our case this is $G^{(1)}$.
This enhanced effect of the sampling error makes the time-series analysis unstable for large $K$.
We can analyze how this weighting alters the previous variance analysis when $\neig=1$.
If we take this into account (see derivation in \ref{app:calc_variance_full}), we find that 
\begin{equation}
\mathrm{Var}(\phi)\propto\frac{1}{KN},
\end{equation}
for a time-series analysis applied to multi-round $k=1$ experiments.

\subsubsection{Classical computation cost}

In practice, the time-series analysis can be split into three calculations; (1) estimation of $\Pfuncbold$ or 
$\mathbb{P}(\frak{m},\frak{n}|\boldsymbol{\phi},\mathbf{A})$, (2) calculation of $g(k)$ from these probabilities via Eq.~(\ref{eq:gk_from_Pfunc}) or Eq.~(\ref{eq:exp-express}), and (3) estimation of the phases $\phi$ from $g(k)$.
Clearly (2) and (3) only need to be done once for the entire set of experiments.

The estimation of the phases $\phi$ requires solving two least squares equations, with cost $O(l^2K)$ (recalling that $l$ is the number of frequencies to estimate, and $K$ is the maximum known value of $g(k)$), and diagonalizing the time-shift matrix $\frak{T}$ with cost $O(l^3)$.
For single-round phase estimation this is the dominant calculation, as calculating $g(k)$ from Eq.~(\ref{eq:gk_from_Pfunc}) requires simply $K$ additions.
As a result this estimator proves to be incredibly fast, able to estimate one frequency from a set of $N=10^6$ experiments of up to $K=10000$ in $<100~\mathrm{ms}$, and $l=1000$ frequencies from $N=10^6$ experiments with $K=1000$ in $<1~\mathrm{min}$.
However, for multi-round phase estimation the calculation of $g(k)$ in Eq.~(\ref{eq:exp-express}) scales as $O(K^4)$.
This then dominates the calculation, requiring $30~\mathrm{s}$ to calculate $50$ points of $g(k)$.
(All calculations performed on a $2.4~\mathrm{GHz}$ Intel i3 processor.)
We note that all the above times are small fractions of the time required to generate the experimental data when $N \gg K$, making this a very practical estimator for near-term experiments.

\subsection{Efficient Bayesian analysis}
\label{sec:bayes}

When the starting state is the eigenstate $|\phi_0\rangle$, the problem of determining $\phi_0$ based on the obtained multi-experiment data has a natural solution via Bayesian methods~\cite{Wie16,BabbushPatent}.
Here we extend such Bayesian methodology to a general starting state. 
For computational efficiency we store a probability distribution over phases $P(\phi)$ using a Fourier representation of this periodic function $P(\phi)$ (see \ref{App:BayesFourier}).
This technique can also readily be applied to the case of Bayesian phase estimation applied to a single eigenstate.

A clearly information-theoretic optimal Bayesian strategy is to choose the $\boldsymbol{\phi}$ and $\boldsymbol{A}$ based on the data obtained in some $N$ experiments \cite{Svo13}. After these $N$ experiments, leading to qubit measurement outcomes $\{{\bf m}_i\}_{i=1}^{N}$, one can simply choose $\bf{A},\boldsymbol{\phi}$ which maximizes the posterior distribution:
\begin{equation}
P_{\rm post}(\boldsymbol{\phi},\mathbf{A})=\frac{P_{\{\mathbf{k}_i\},\{\boldsymbol{\beta}_i\}}(\{\mathbf{m}_i\}|\boldsymbol{\phi},\mathbf{A})}{P(\{\mathbf{m}_i\})}P_{\rm prior}(\boldsymbol{\phi},\mathbf{A}),
\end{equation}
In other words, one chooses
\begin{align*}
&(\boldsymbol{\phi}_{\rm opt}, \mathbf{A}_{\rm opt})=\arg \max_{\boldsymbol{\phi},\mathbf{A}}  \log 
P_{\rm post}(\boldsymbol{\phi},\mathbf{A}) \\
&=\arg \max_{\boldsymbol{\phi},\mathbf{A}} \left[\log P_{\{\mathbf{k}_i\},\{\boldsymbol{\beta}_i\}}(\{\mathbf{m}_i\}|\boldsymbol{\phi},\mathbf{A})+\log P_{\rm prior}(\boldsymbol{\phi},\mathbf{A})\right].
\end{align*}
A possible way of implementing this strategy is to (1) assume the prior distribution to be independent of $\mathbf{A}$ and $\boldsymbol{\phi}$ and (2) estimate the maximum by assuming that the derivative with respect to $\mathbf{A}$ and $\boldsymbol{\phi}$ vanishes at this maximum.

Instead of this method we update our probability distribution over $\boldsymbol{\phi}$ and $\mathbf{A}$ after each experiment. After experiment $n$ the posterior distribution $P_n(\boldsymbol{\phi},\mathbf{A})$ via Bayes' rule reads
\begin{equation}
P_n(\boldsymbol{\phi},\mathbf{A})=\frac{P_{\mathbf{k},\boldsymbol{\beta}}(\mathbf{m}|\boldsymbol{\phi},\mathbf{A})}{P(\mathbf{m})}P_{n-1}(\boldsymbol{\phi},\mathbf{A})\label{eq:Bayesrule}.
\end{equation}
To calculate the updates we will assume that the distribution over the phases $\phi_j$ and probabilities $A_j$ are independent, that is,
\begin{equation}
P_n(\boldsymbol{\phi},\mathbf{A})=P_n^{\rm red}(\mathbf{A})\prod_{j=0}^{\neig-1}P_n^{j}(\phi_j)\label{eq:Indep_distribution}.
\end{equation}

As prior distribution we take $P_0(\boldsymbol{\phi},\mathbf{A})=P_{\rm prior}(\mathbf{A})P_{\rm prior}(\boldsymbol{\phi})$ with a flat prior $P_{\rm prior}(\boldsymbol{\phi})=(\frac{1}{2\pi})^{\neig}$, given the absence of a more informed choice.
We take $P_{\rm prior}(\mathbf{A})=e^{-(\mathbf{A}-\mathbf{A}_0)^2/2\Sigma^2}$, with $\mathbf{A}_0$ and $\Sigma^2$ approximate mean and covariance matrices.
We need to do this to break the symmetry of the problem, so that $\tilde{\phi}_0 $ is estimating $\phi_0$ and not any of the other $\phi$s.
We numerically find that the estimator convergence is relatively independent of our choice of $\mathbf{A}_0$ and $\Sigma^2$.

The approximation in Eq.~(\ref{eq:Indep_distribution}) allows for relatively fast calculations of the Bayesian update of $P_n^j(\phi_j)$, and an approximation to the maximum-likelihood estimation of $P_n^{\rm red}(\mathbf{A})$.
Details of this computational implementation are given in \ref{App:BayesFourier-multi}.

\subsubsection{Classical computation cost}
In contrast to the time-series estimator, the Bayesian estimator incurs a computational cost in processing the data from each individual experiment. On the other hand, obtaining the estimate $\tilde{\phi}_0$ for $\phi_0$ is simple, once one has the probability distribution $P^{j=0}(\phi)$:
\begin{align*}
\tilde{\phi}_0=\mathrm{arg}(\int d\phi P^{j=0}(\phi) e^{i\phi}).
\end{align*}

A key parameter here is the number of frequencies $\#{\rm freq}$ stored in the Fourier representation of $P(\phi)$; each update requires multiplying a vector of length $\nfreq$ by a sparse matrix.
Our approximation scheme for calculating the update to $\mathbf{A}$ makes this multiplication the dominant time cost of the estimation.
As we argue in \ref{App:BayesFourier-multi} one requires $\#{\rm freq}\geq K_{\rm tot}$ to store a fully accurate representation of the probability vector. For the single-round scenario with $k_r=1$, hence $K_{\rm tot}=N$, we find a large truncation error when $\#{\rm freq}\ll N$, and so the computation cost scales as $N^2$. In practice we find that processing the data from $N<10^4$ experiments takes seconds on a classical computer, but processing more than $10^5$ experiments becomes rapidly unfeasible.

\subsection{Experiment design}\label{sec:expt_design}

Based on the considerations above we seek to compare some choices for the meta-parameters in each experiment, namely the number of rounds, and the input parameters $k_r$ and $\beta_r$ for each round. 

Previous work~\cite{Wie14,Wie16}, which took as a starting state the eigenstate $|\phi_0\>$, formulated a choice of $k$ and $\beta$, using single-round experiments and Bayesian processing, namely
\begin{equation}
k=\min\left(\left\lceil\frac{1.25}{\sigma_{P_n^{j=0}(\phi_0)}}\right\rceil,K_{\rm err} \right),\hspace{0.5cm}\beta \sim P_{n}^{j=0}(\phi_0=\beta)\label{eq:metaparams},
\end{equation}
Roughly, this heuristic adapts to the expected noise in the circuit by not using any $k$ such that the implementation of $U^k$ takes longer than $T_{\rm err}/\nsys$.
It also adapts $k$ to the standard-deviation of the current posterior probability distribution over $\phi_0$: a small standard-deviation after the $n$th experiment implies that $k$ should be chosen large to resolve the remaining bits in the binary expansion of $\phi_0$~\footnote{Note that this strategy is the opposite of textbook phase estimation in which one necessarily learns the least-significant bit of $\phi_0$ {\em first} by choosing the largest $k$. One chooses the next smallest $k$ and $\beta$ so that the next measurement outcome gives the next more-significant bit etc.}.

In this work we use a starting state which is not an eigenstate, and as such we must adjust the choice in Eq.~(\ref{eq:metaparams}).
As noted in Sec.~\ref{sec:data_analysis}, to separate different frequency contributions to $g(k)$ we need good accuracy beyond that at a single value of $k$.
The optimal choice of the number of frequencies to estimate depends on the distribution of the $A_j$, which may not be well known in advance.
Following the inspiration of~\cite{Wie16}, we choose for the Bayesian estimator
\begin{align}
k & \in \{1,\ldots,K\} \nonumber \\
K & = \min\left(\left\lceil\frac{1.25}{\sigma_{P_n^{j=0}(\phi_0)}}\right\rceil,K_{\rm err} \right).\label{eq:k_choice}
\end{align}
We thus similarly bound $K$ depending how well one has already converged to a value for $\phi_0$ which constitutes some saving of resources.
At large $N$ we numerically find little difference between choosing $k$ at random from $\{1,\ldots, K\}$ and cycling through $k=1,\ldots,K$ in order.
For this Bayesian estimator we draw $\beta$ at random from a uniform distribution $[0,2\pi)$.
We find that the choice of $\beta$ has no effect on the final estimation (as long as it is not chosen to be a single number) 
For the time-series estimator applied to single-round experiments, we choose to cycle over $k=1,\ldots,K$ so that it obtains a complete estimate of $g(k)$ as soon as possible, taking an equal number of experiments with final rotation $\beta=0$ and $\beta=\pi/2$ at each $k$.
Here again $K \leq K_{\rm err}$, so that we choose the same number of experiments for each $k \leq K$.
For the time-series estimator applied to multi-round experiments, we choose an equal number of rounds with $\beta=0$ and $\beta=\pi/2$, taking the total number of rounds equal to $R=K$.

\begin{figure}
\includegraphics[width=0.8\columnwidth]{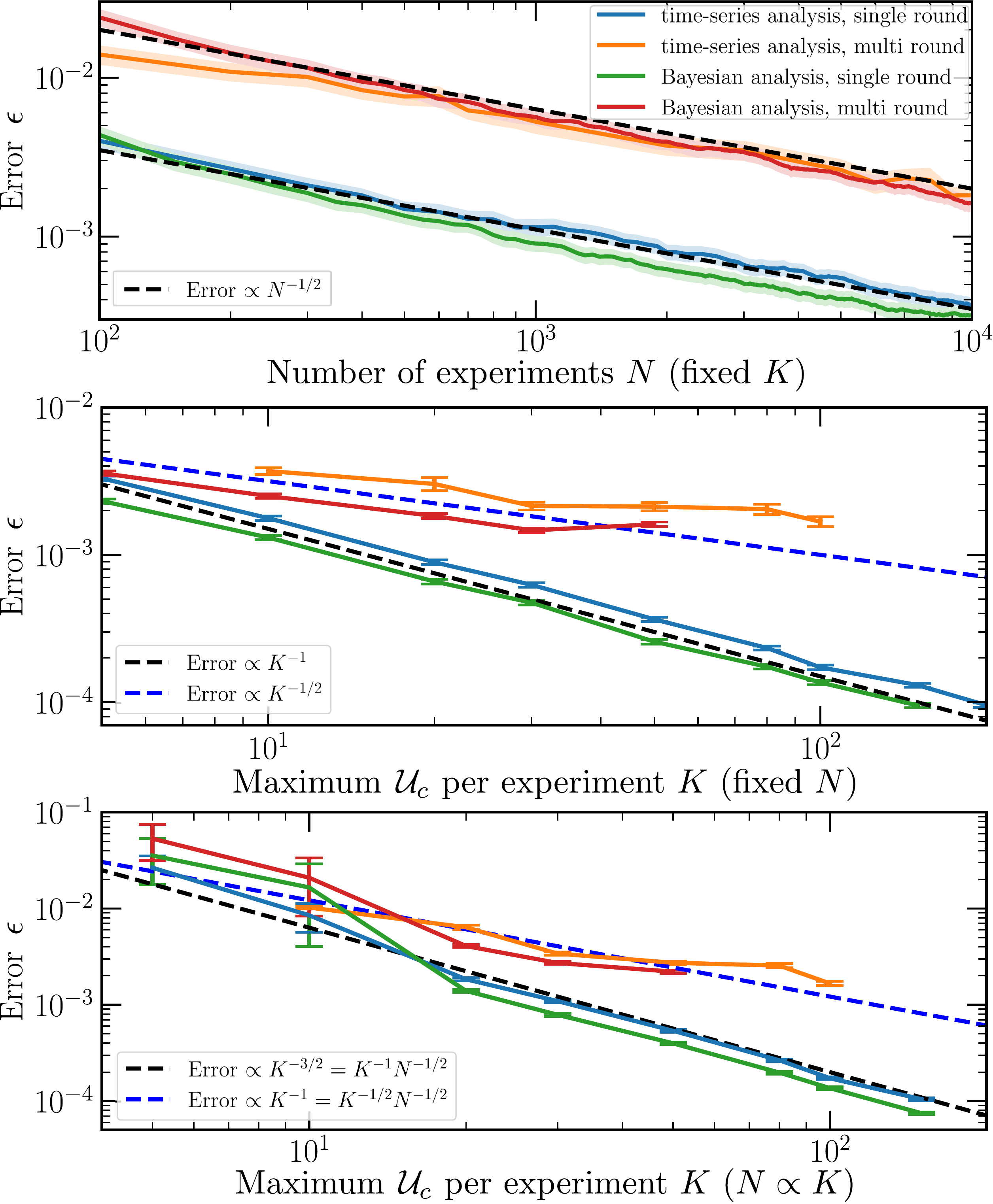}
\caption{\label{fig:scaling}Estimator performance for single eigenvalues with single and multi-round $k=1$ QPE schemes. Plots show scaling of the mean absolute error (Eq.~(\ref{eq:epsilon})) with (top) the number of experiments (at fixed $K=50$), with (middle) $K$ for a fixed total number of experiments ($N=10^6$), and (bottom) with $K$ with a fixed number ($100$) of experiments per $k=1,\ldots,K$ (i.e. $N=200K$). Data is averaged over $200$-$500$ QPE simulations, with a new eigenvalue chosen for each simulation. Shaded regions (top) and error bars (middle, bottom) give $95\%$ confidence intervals. Dashed lines show the scaling laws of Eq.~(\ref{eq:time_series_var}) (fitted by eye). The top-right legend labeling the different estimation schemes is valid for all three plots.}
\end{figure}

\section{Results without experimental noise}
\label{sec:num}

We first focus on the performance of our estimators in the absence of experimental noise, to compare their relative performance and check the analytic predictions in Sec.~\ref{sec:time}.
Although with a noiseless experiment our limit for $K$ is technically infinite, we limit it to a make connection with the noisy results of the following section.
Throughout this section we generate results directly by calculating the function $\Pfuncbold$ and sampling from it. Note that $\Pfuncbold$ only depends on $\neig$ and not on the number of qubits in the system.

\subsection{Single eigenvalues}\label{sec:1ev}
To confirm that our estimators achieve the scaling bounds discussed previously, we first test them on the single eigenvalue scenario $\neig=1$.
In Fig.~\ref{fig:scaling}, we plot the scaling of the average absolute error in an estimation $\tilde{\phi}$ of a single eigenvalue $\phi\in[-\pi,\pi)$, defined so as to respect the $2\pi$-periodicity of the phase:
\begin{equation}
\epsilon:=\left\<\min\left(|\phi-\tilde{\phi}|,2\pi-|\phi-\tilde{\phi}|\right)\right\>=\left\<\left|\mathrm{Arg}\left(\vphantom{|\phi-\tilde{\phi}|,2\pi-|\phi-\tilde{\phi}|}e^{i(\phi-\tilde{\phi})}\right)\right|\right\>,\label{eq:epsilon}
\end{equation}
as a function of varying $N$ and $K$. Here $\<\>$ represents an average over repeated QPE simulations, and the $\mathrm{Arg}$ function is defined using the range $[-\pi,\pi)$ (otherwise the equality does not hold).

We see that both estimators achieve the previously-derived bounds in \ref{sec:time} (overlayed as dashed lines), and both estimators achieve almost identical convergence rates. The results for the Bayesian estimation match the scaling observed in Ref.~\cite{Wie16}.
Due to the worse scaling in $K$, the multi-round $k=1$ estimation significantly underperforms single-round phase estimation.
This is a key observation of this paper, showing that if the goal is to estimate a phase rather than to project onto an eigenstate, it is preferable to do single-round experiments.

\subsection{Example behaviour with multiple eigenvalues}\label{sec:single_shot}

\begin{figure}
\includegraphics[width=0.8\columnwidth]{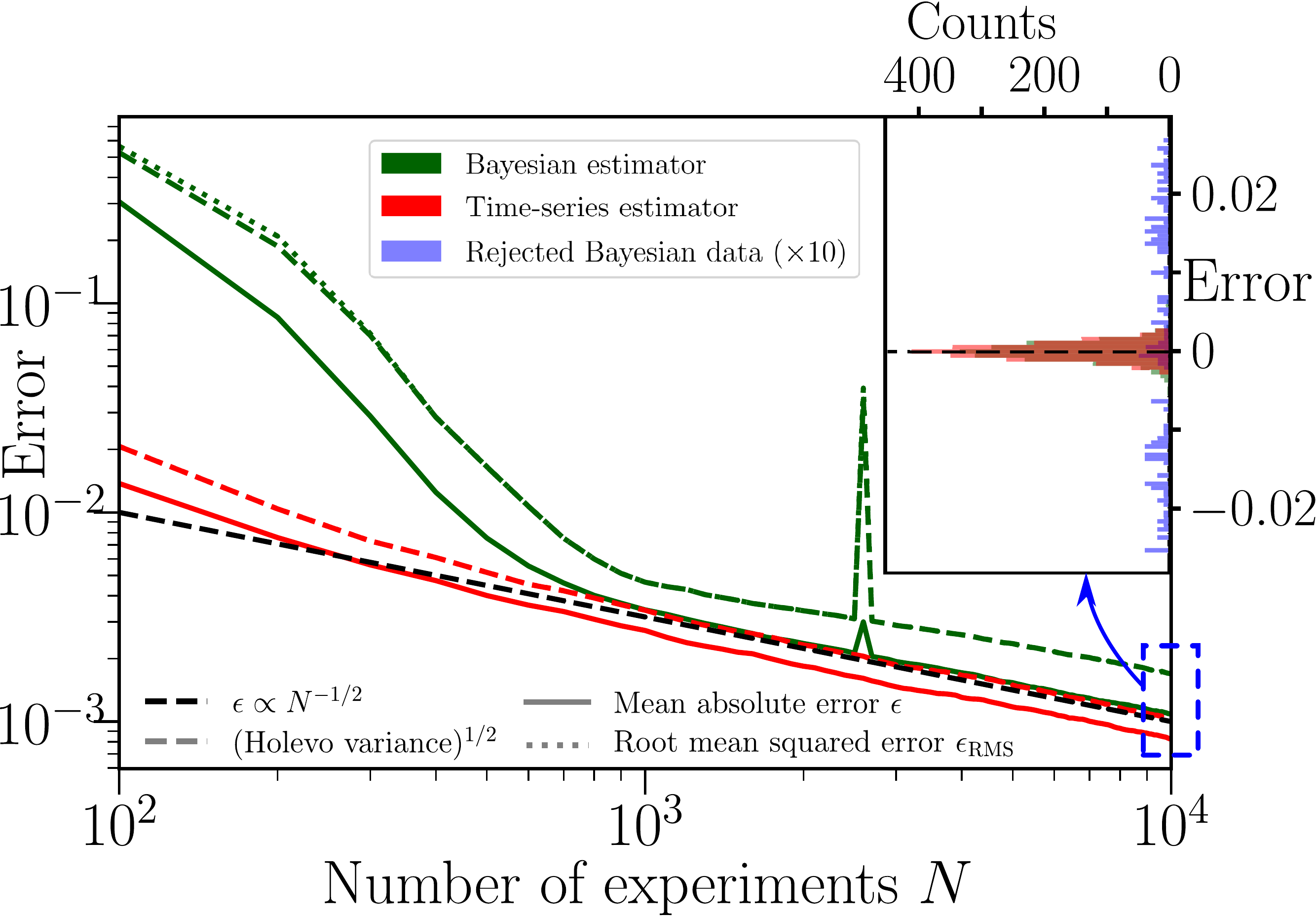}
\caption{\label{fig:single_shot}Scaling of error for time-series (dark green) and Bayesian (red) estimators with the number of experiments performed for a single shot of a unitary with randomly drawn eigenphases (parameters given in text). Three error metrics are used as marked (described in text - note that the mean squared error and Holevo variance completely overlap for the time-series estimator). Data is averaged over $2000$ simulations. The peak near $N=3000$ comes from deviation in a single simulation and is not of particular interest. With this exception, error bars are approximately equal to width of the lines used. (Inset) histogram of the estimated phases after $N=10^4$ experiments. Blue bars correspond to Bayesian estimates that were rejected (rejection method described in text). These have been magnified $10\times$ to be made visible.}
\end{figure}

The performance of quantum phase estimation is dependent on both the estimation technique and the system being estimated.
Before studying the system dependence, we first demonstrate that our estimators continue to perform at all in the presence of multiple eigenvalues.
In Fig.~\ref{fig:single_shot}, we demonstrate the convergence of both the Bayesian and time-series estimators in the estimation of a single eigenvalue $\phi_0=-0.5$ of a fixed unitary $U$, given a starting state $|\Psi_0\>$ which is a linear combination of $10$ eigenstates $|\phi_j\>$.
We fix $|\<\phi_0|\Psi_0\>|^2=0.5$, and draw other eigenvalues and amplitudes at random from $[0,\pi]$ (making the minimium gap $\phi_j-\phi_0$ equal to $0.5$).
We perform $2000$ QPE simulations with $K=50$, and calculate the mean absolute error $\epsilon$ (Eq.~(\ref{eq:epsilon}), solid), Holevo variance $\left|\left\<e^{i\tilde{\phi}}\right\>\right|^{-2}-1$ (dashed), and root mean squared error $\epsilon_{\mathrm{RMS}}$ (dotted), given by
\begin{equation}
\epsilon^2_{\mathrm{RMS}}:=\left\<\min\left(|\phi-\tilde{\phi}|,2\pi-|\phi-\tilde{\phi}|\right)^2\right\>=\left\<\left|\mathrm{Arg}\left(\vphantom{|\phi-\tilde{\phi}|,2\pi-|\phi-\tilde{\phi}|}e^{i(\phi-\tilde{\phi})}\right)\right|^2\right\>\label{eq:epsilon}.
\end{equation}
We observe that both estimators retain their expected $\epsilon\propto N^{-1/2}$, with one important exception.
The Bayesian estimator occasionally ($10\%$ of simulations) estimates multiple eigenvalues near $\phi_0$.
When this occurs, the estimations tend to repulse each other, making neither a good estimation of the target.
This is easily diagnosable without knowledge of the true value of $\phi_0$ by inspecting the gap between estimated eigenvalues.
While using this data to improve estimation is a clear target for future research, for now we have opted to reject simulations where such clustering occurs (in particular, we have rejected datapoints where $\min(\bar{\phi}_0-\bar{\phi}_j) < 0.05$).
That this is required is entirely system-dependent: we find the physical Hamiltonians studied later in this text to not experience this effect.
We attribute this difference to the distribution of the amplitudes $A_j$ - physical Hamiltonians tend to have a few large $A_j$, whilst in this simulation the $A_j$ were distributed uniformly.

In the inset to Fig.~\ref{fig:single_shot}, we plot a histogram of the estimated eigenphases after $N=10^4$ experiments.
For the Bayesian estimator, we show both the selected (green) and rejected (blue) eigenphases.
We see that regardless of whether rejection is used, the distribution appears symmetric about the target phase $\phi_0$.
This suggests that in the absence of experimental noise, both estimators are unbiased.
Proving this definitively for any class of systems is difficult, but we expect both estimators to be unbiased provided $A_0\gg1/K$.
When $A_0\leq 1/K$, one can easily construct systems for which no phase estimation can provide an unbiased estimation of $\phi_0$ (following the arguments of Sec.~\ref{sec:data_analysis}).
We further see that the scaling of the RMS error $\epsilon_{\mathrm{RMS}}$ and the Holevo variance match the behaviour of the mean absolute error $\epsilon$, implying that our results are not biased by the choice of estimator used.

\subsection{Estimator scaling with two eigenvalues}\label{sec:2ev}

\begin{figure*}[htb]
\includegraphics[width=0.8\textwidth]{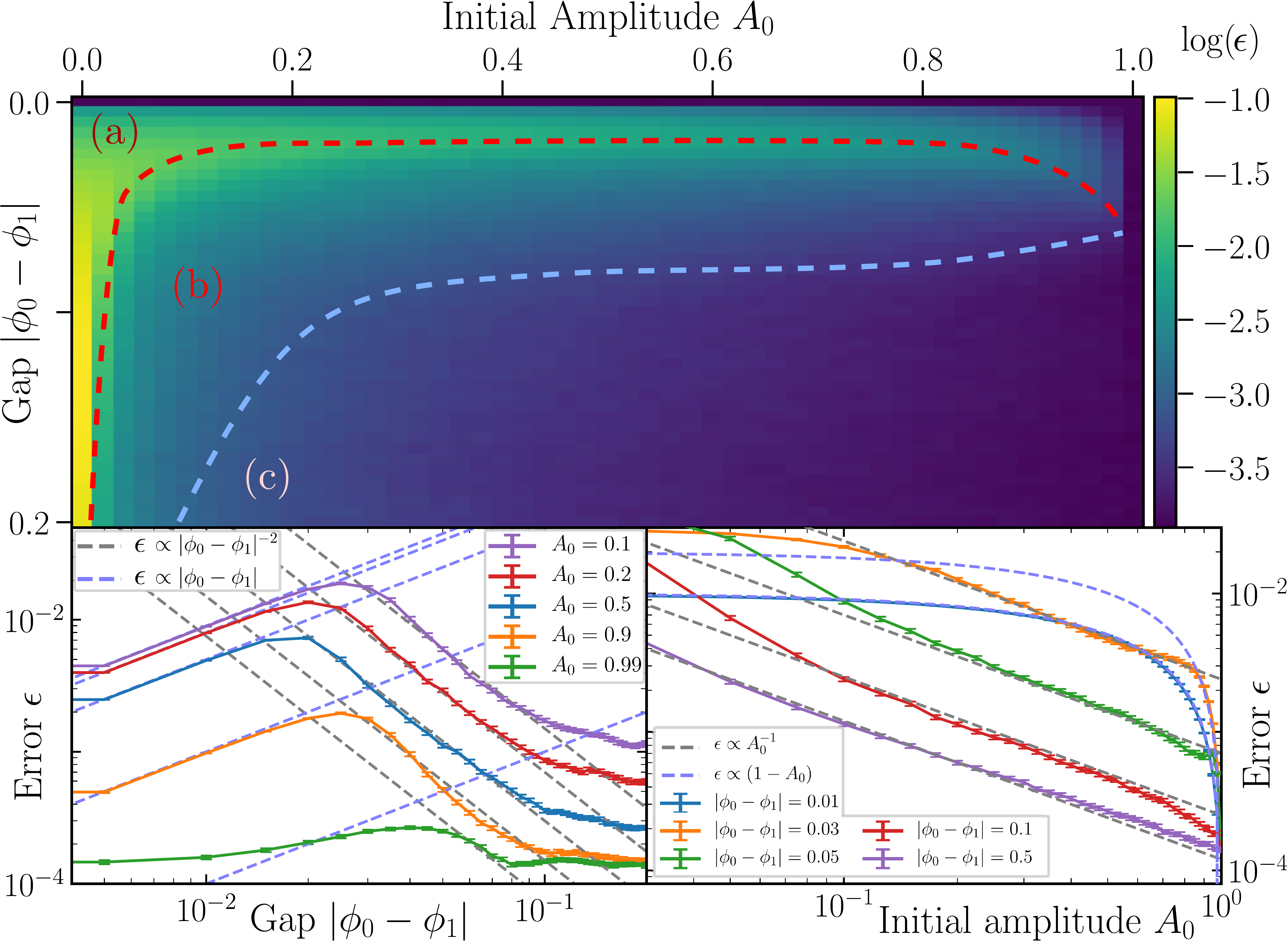}
\caption{\label{fig:2ev}Performance of the time-series estimator in the presence of two eigenvalues. (top) Surface plot of the error after $N=10^6$ experiments for $K=50$, as a function of the overlap $A_0$ with the target state $|\phi_0\>$, and the gap $|\phi_0-\phi_1|$. Plot is divided by hand into three labeled regions where different scaling laws are observed. Each point is averaged over $500$ QPE simulations. (bottom) log-log plots of vertical (bottom left) and horizontal (bottom right) cuts through the surface, at the labeled positions. Dashed lines in both plots are fits (by eye) to the observed scaling laws. Each point is averaged over $2000$ QPE simulations, and error bars give $95\%$ confidence intervals.}
\end{figure*}

The ability of QPE to resolve separate eigenvalues at small $K$ can be tested in a simple scenario of two eigenvalues, $\phi_0$ and $\phi_1$. The input to the QPE procedure is then entirely characterized by the overlap $A_0$ with the target state $|\phi_0\>$, and the gap $\delta=|\phi_0-\phi_1|$.

In Fig.~\ref{fig:2ev}, we study the performance of our time-series estimator in estimating $\phi_0$ after $N=10^6$ experiments with $K=50$, measured again by the mean error $\epsilon$ (Eq.~(\ref{eq:epsilon})).
We show a two-dimensional plot (averaged over $500$ simulations at each point $A_0,\delta$) and log-log plots of one-dimensional vertical (lower left) and horizontal (lower right) cuts through this surface.
Due to computational costs, we are unable to perform this analysis with the Bayesian estimator, or for the multi-round scenario.
We expect the Bayesian estimator to have similar performance to the time-series estimator (given their close comparison in Sec.~\ref{sec:1ev} and Sec.~\ref{sec:single_shot}).
We also expect the error in multi-round QPE to follow similar scaling laws in $A_0$ and $\delta$ as single-round QPE (i.e.~multi-round QPE should be suboptimal only in its scaling in $K$).

The ability of our estimator to estimate $\phi_0$ in the presence of two eigenvalues can be split into three regions (marked as $(a)$, $(b)$, $(c)$ on the surface plot).
In region $(a)$, we have performed insufficient sampling to resolve the eigenvalues $\phi_0$ and $\phi_1$, and QPE instead estimates the weighted average phase $A_0\phi_0+A_1\phi_1$.
The error in the estimation of $\phi_0$ then scales by how far it is from the average, and how well the average is resolved
\begin{equation}
\epsilon\propto (1-A_0)\delta K^{-1}N^{-1/2}.
\end{equation}
In region $(b)$, we begin to separate $\phi_0$, from the unwanted frequency $\phi_1$, and our convergence halts,
\begin{equation}
\epsilon\propto A_0^{-1}\delta^{-2}.
\end{equation}
In region $(c)$, the gap is sufficiently well resolved and our estimation returns to scaling well with $N$ and $K$
\begin{equation}
\epsilon\propto A_0^{-1}K^{-1}N^{-1/2}.
\end{equation}
The scaling laws in all three regions can be observed in the various cuts in the lower plots of Fig.~\ref{fig:2ev}.
We note that the transition between the three regions is not sharp (boundaries estimated by hand), and is $K$ and $N$-dependent.

\subsection{Many eigenvalues}

\begin{figure}[htb]
\includegraphics[width=0.8\columnwidth]{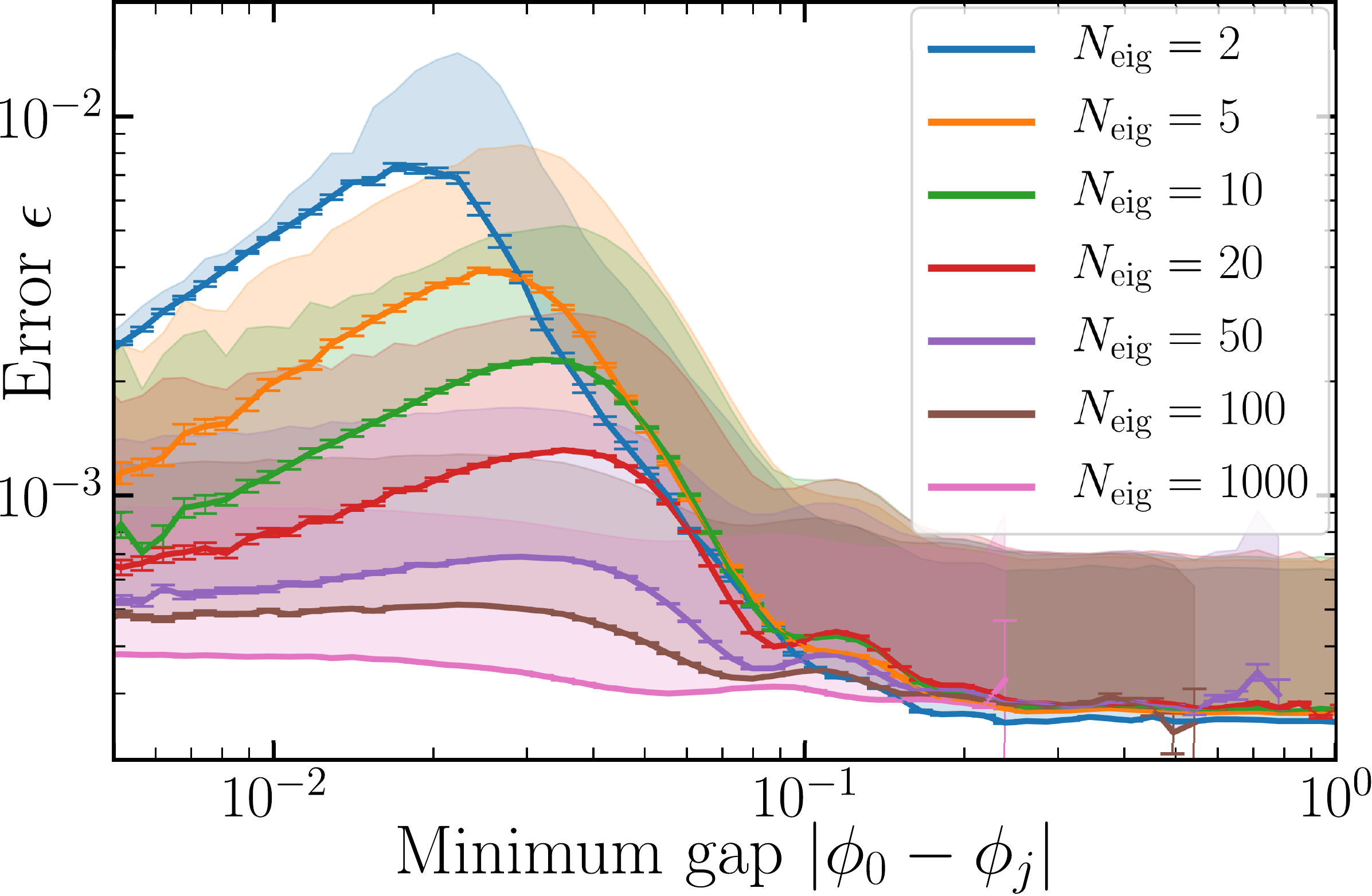}
\caption{\label{fig:many_ev}Performance of the time-series estimator in the presence of multiple eigenvalues. Error bars show $95\%$ confidence intervals (data points binned from $4\times 10^6$ simulations). Shaded regions show upper $2\sigma$ interval of data for each bin.}
\end{figure}

To show that our observed scaling is applicable beyond the toy $2$-eigenvalue system, we now shift to studying systems of random eigenvalues with $\neig>1$.
In keeping with our insight from the previous section, in Fig.~\ref{fig:many_ev} we fix $\phi_0=0$, and study the error $\epsilon$ as a function of the gap 
\begin{equation}
\delta=\min_{j>1}(|\phi_j-\phi_0|).
\end{equation}
We fix $A_0=0.5$, and draw the other parameters for the system from a uniform distribution: $\phi_j\sim[\delta,\pi]$, $A_j\sim[0,0.5]$ (fixing $\sum_{j=1}^{\neig}A_j=1-A_0$).
We plot both the average error $\epsilon$ (line) and the upper $47.5\%$ confidence interval $[\epsilon,\epsilon+2\sigma_\epsilon]$ (shaded region) for various choices of $\neig$.
We observe that increasing the number of spurious eigenvalues does not critically affect the error in estimation; indeed the error generally decreases as a function of the number of eigenvalues.
This makes sense; at large $\neig$ the majority of eigenvalues sit in region $(c)$ of Fig.~\ref{fig:2ev}, and we do not expect these to combine to distort the estimation.
Then, the nearest eigenvalue $\min_{j\neq 0}\phi_j$ has on average an overlap $A_j\propto 1/\neig$, and its average contribution to the error in estimating $\phi_0$ (inasmuch as this can be split into contributions) scales accordingly.
We further note that the worst-case error remains that of two eigenvalues at the crossover between regions $(a)$ and $(b)$.
In \ref{app:multi_evdata} we study the effect of confining the spurious eigenvalues to a region $[\delta,\phi_{\max}]$.
We observe that when most eigenvalues are confined to regions $(a)$ and $(b)$, the scaling laws observed in the previous section break down, however the worst-case behaviour remains that of a single spurious eigenvalue.
This implies that sufficiently long $K$ is not a requirement for QPE, even in the presence of large systems or small gaps $\delta$; it can be substituted by sufficient repetition of experiments.
However, we do require that the ground state is guaranteed to have sufficient overlap with the starting state - $A_0>1/K$ (as argued in Sec.~\ref{sec:data_analysis}).
As QPE performance scales better with $K$ than it does with $N$, a quantum computer with coherence time $2T$ is still preferable to two quantum computers with coherence time $T$ (assuming no coherent link between the two).

\section{The effect of experimental noise}\label{sec:noise}

Experimental noise currently poses the largest impediment to useful computation on current quantum devices.
As we suggested before, experimental noise limits $K$ so that for $K\gtrsim K_{\rm err}$ the circuit is unlikely to produce reliable results.
However, noise on quantum devices comes in various flavours, which can have different corrupting effects on the computation.
Some of these corrupting effects (in particular, systematic errors) may be compensated for with good knowledge of the noise model.
For example, if we knew that our system applied $U=e^{-i \Hh(t+\epsilon)}$ instead of $U=e^{-i\Hh t}$, one could divide $\tilde{\phi_0}$ by $(t+\epsilon)/t$ to precisely cancel out this effect.
In this study we have limited ourselves to studying and attempting to correct two types of noise: depolarizing noise, and circuit-level simulations of superconducting qubits.
Given the different effects observed, extending our results to other noise channels is a clear direction for future research.
In this section we do not study multi-round QPE, so each experiment consists of a single round.
A clear advantage of the single-round method is that the only {\em relevant} effect of any noise in a single-round experiment is to change the outcome of the ancilla qubit, independent of the number of system qubits $\nsys$.

\subsection{Depolarizing noise}

\begin{figure}[htb]
\includegraphics[width=0.8\columnwidth]{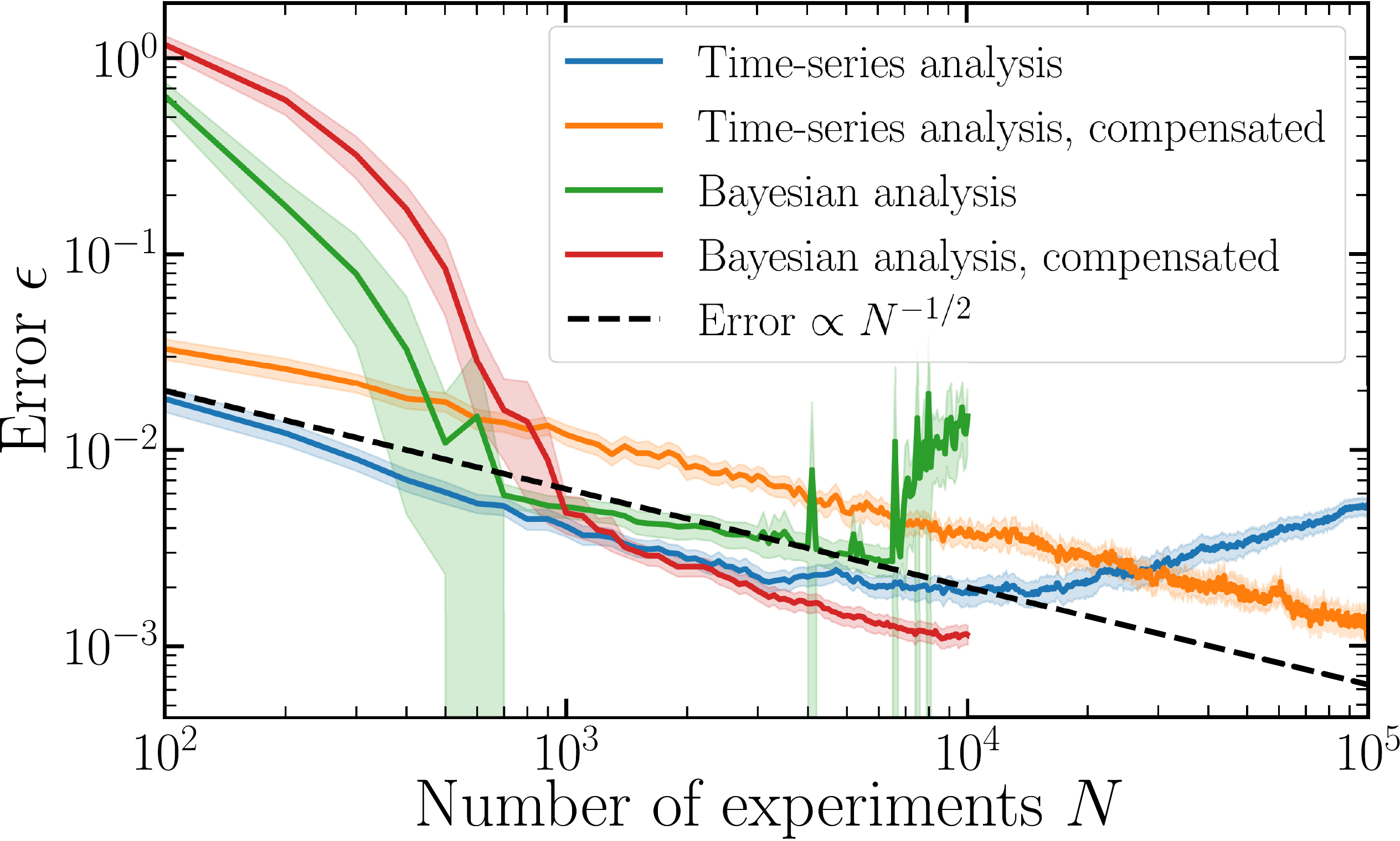}
\caption{\label{fig:depol_noise}Convergence of Bayesian and time-series estimators in the presence of depolarizing noise and multiple eigenvalues, both with and without noise compensation techniques (described in text). Fixed parameters for all plots are given in text. Shaded regions denote a $95\%$ confidence interval (data estimated over $200$ QPE simulations). The black dashed line shows the $N^{-1/2}$ convergence expected in the absence of sampling noise. Data for the Bayesian estimator was not obtained beyond $N=10^4$ due to computational constraints.}
\end{figure}

A very simple noise model is that of depolarizing noise, where the outcome of each experiment is either correct with some probability $p$ or gives a completely random bit with probability $1-p$.
We expect this probability $p$ to depend on the circuit time and thus the choice of $k\geq 0$, i.e.
\begin{equation}
p=p(k)=e^{-k/K_{\rm err}}.
\label{eq:kdep}
\end{equation}
We can simulate this noise by directly applying it to the calculated probabilities $\Pfuncsr$ for a single round 
\begin{equation}
\Pfuncsr\rightarrow\Pfuncsr p(k)+\frac{1-p(k)}{2}\label{eq:Pfunc_adjust}.
\end{equation}
In Fig.~\ref{fig:depol_noise}, we plot the convergence of the time-series (blue) and Bayesian (green) estimators as used in the previous section as a function of the number of experiments, with fixed $K=50=K_{\rm err}/2$ fixed, $A_0=0.5$, $\neig=10$ and $\delta=0.5$.
We see that both estimators obey $N^{-1/2}$ scaling for some portion of the experiment, however this convergence is unstable, and stops beyond some critical point.

Both the Bayesian and time-series estimator can be adapted rather easily to compensate for this depolarizing channel.
To adapt the time-series analysis, we note that the effect of depolarizing noise is to send $g(k) \rightarrow g(k)p(k)$ when $k>0$, via Eq.~(\ref{eq:gk_from_Pfunc}) and Eq.~(\ref{eq:Pfunc_adjust}).
Our time-series analysis was previously performed over the range $k=-K,\ldots,K$ (getting $g(-k)=g^*(k)$ for free), and over this range
\begin{equation}
g(k)\rightarrow g(k)p(|k|).
\end{equation}
$g(k)$ is no longer a sum of exponential functions over our interval $[-K,K]$, as it is not differentiable at $k=0$, which is the reason for the failure of our time-series analysis.
However, over the interval $[0,K]$ this is not an issue, and the time-series analysis may still be performed.
If we construct a shift operator $T$ using $g(k)$ from $k=0,\ldots,K$, this operator will have eigenvalues $e^{i \phi_j-1/K_{\rm err}}$.
This then implies that the translation operator $T$ can be calculated using $g(k)$ with $k>0$, and the complex argument of the eigenvalues of $T$ give the correct phases $\phi_j$.
We see that this is indeed the case in Fig.~\ref{fig:depol_noise} (orange line).
Halving the range of $g(k)$ that we use to estimate $\phi_0$ decreases the estimator performance by a constant factor, but this can be compensated for by increasing $N$.

Adapting the Bayesian estimator requires simply that we use the correct conditional probability, Eq.~(\ref{eq:Pfunc_adjust}).
This in turn requires that we either have prior knowledge of the error rate $K_{\rm err}$, or estimate it alongside the phases $\phi_j$.
For simplicity, we opt to choose the former.
In an experiment $K_{\rm err}$ can be estimated via standard QCVV techniques, and we do not observe significant changes in estimator performance when it is detuned.
Our Fourier representation of the probability distribution of $\phi_0$ can be easily adjusted to this change.
The results obtained using this compensation are shown in Fig.~\ref{fig:depol_noise}: we observe that the data follows a $N^{-1/2}$ scaling again.

\subsection{Realistic circuit-level noise}\label{sec:realistic_noise}
Errors in real quantum computers occur at a circuit-level, where individual gates or qubits get corrupted via various error channels.
To make connection to current experiments, we investigate our estimation performance on an error model of superconducting qubits.
Full simulation details can be found in \ref{app:sim_details}.
Our error model is primarily dominated by $T_1$ and $T_2$ decoherence, incoherent two-qubit flux noise, and dephasing during single-qubit gates.
We treat the decoherence time $T_{\rm err}=T_1=T_2$ as a free scale parameter to adjust throughout our simulations, whilst keeping all other error parameters tied to this single scale parameter for simplicity.
In order to apply circuit-level noise we must run quantum circuit simulations, for which we use the quantumsim density matrix simulator first introduced in~\cite{Obr17}.
We then choose to simulate estimating the ground state energy of four hydrogen atoms in varying rectangular geometries, with Hamiltonian $\Hh$ taken in the STO-3G basis calculated via psi4~\cite{Par17}, requiring $\nsys=8$ qubits.
We make this estimation via a lowest-order Suzuki-Trotter approximation~\cite{Suz76} to the time-evolution operator $e^{-i\Hh t}$.
To prevent energy eigenvalues wrapping around the circle we fix $t=1/\sqrt{\mathrm{Trace}[\Hh^\dag \Hh]/(2^{\nsys})}$~\footnote{This normalization is not good for large systems since it makes $t$ exponentially small in system size. A scalable choice for normalization is to first determine upper and lower bounds on the eigenvalues of $\Hh$ present in the starting state, assume that they occur in a some numerical window $W$. Given $W$ (which is at most ${\rm poly}(\nsys)$), one sets $U=\exp(-i \pi \Hh/W)$.  The implementation of this $U$ in Trotterized form with sufficient accuracy determines $T_U$.}.
The resultant $9$-qubit circuit is made using the OpenFermion package~\cite{Mcc17}.

In lieu of any circuit optimizations (e.g.~\cite{Ber17,Has15}), the resulting circuit has a temporal length per unitary of $T_U=42~\mu\mathrm{s}$ (with single- (two-) qubit gate times $20~\mathrm{ns}$ ($40~\mathrm{ns}$)).
This makes the circuit unrealistic to operate at current decoherence times for superconducting circuits, and we focus on decoherence times $1-2$ orders of magnitude above what is currently feasible, i.e.~$T_{\rm err}=5-50~\mathrm{ms}$.
However one may anticipate that the ratio $T_U/T_{\rm err}$ can be enlarged by circuit optimization or qubit improvement.
Naturally, choosing a smaller system, less than 8 qubits, or using error mitigation techniques could also be useful.

We observe realistic noise to have a somewhat different effect on both estimators than a depolarizing channel.
Compared to the depolarizing noise, the noise may (1) be biased towards 0 or 1 and/or (2) its dependence on $k$ may not have the form of Eq.~(\ref{eq:kdep}). 

\begin{figure}
\includegraphics[width=0.8\columnwidth]{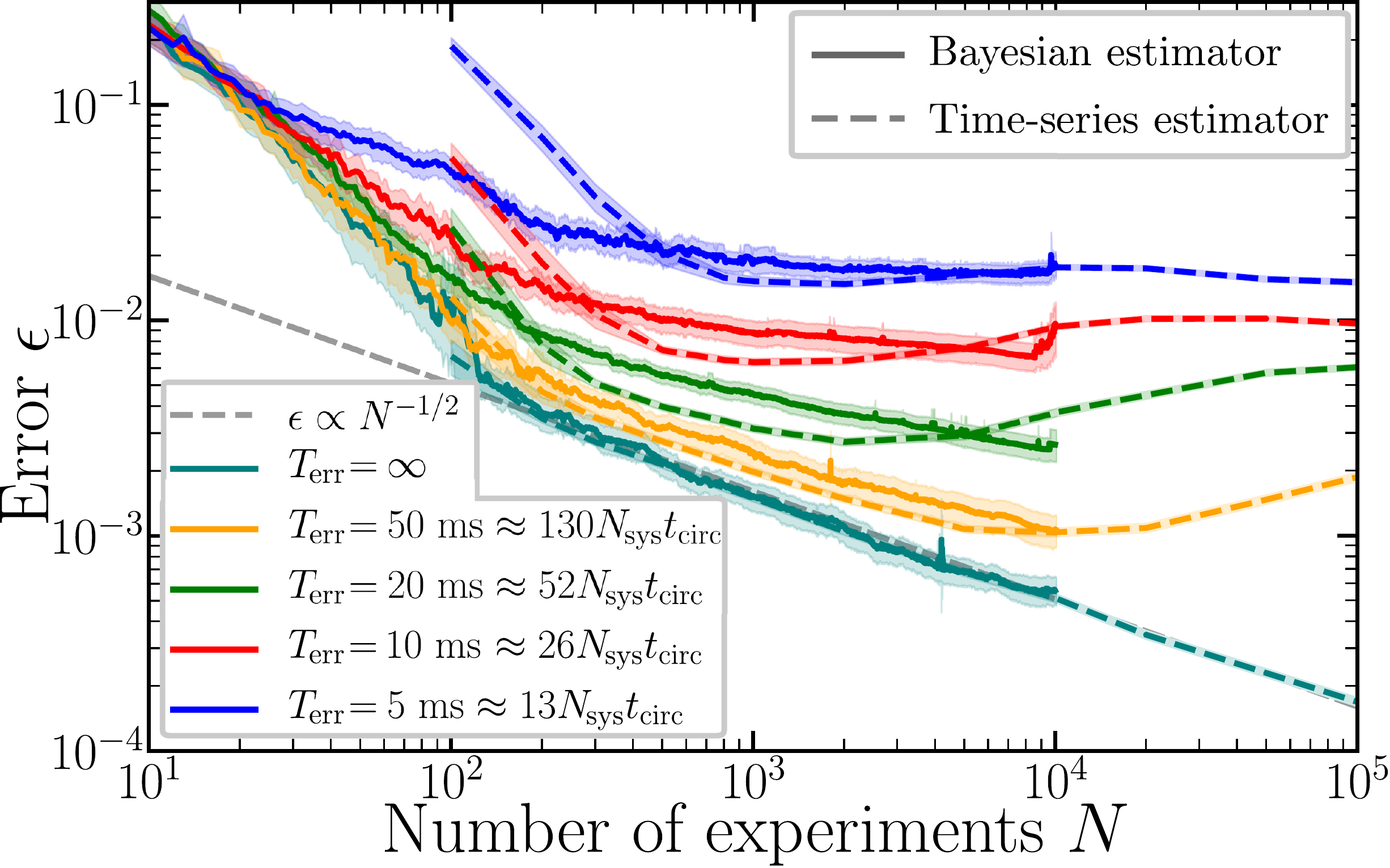}
\caption{\label{fig:noise_vs_nonoise} Performance of Bayesian (solid) and time-series (dashed) estimators in the presence of realistic noise without any compensation techniques. Shaded regions denote $95\%$ confidence intervals (averaged over $100-500$ QPE simulations). The time-series analysis requires $N>2K$ experiments in order to produce an estimate, and so its performance is not plotted for $N<100$.}
\end{figure}

In Fig.~\ref{fig:noise_vs_nonoise}, we plot the performance of both estimators at four different noise levels (and a noiseless simulation to compare), in the absence of any attempts to compensate for the noise.
Unlike for the depolarizing channel, where a $N^{-1/2}$ convergence was observed for some time before the estimator became unstable, here we see both instabilities and a loss of the $N^{-1/2}$ decay to begin with.
Despite this, we note that reasonable convergence (to within $1-2\%$) is achieved, even at relatively low coherence times such as $K_{\rm err}=10$.
Regardless, the lack of eventual convergence to zero error is worrying, and we now shift to investigating how well it can be improved for either estimator.

\begin{figure}
\includegraphics[width=0.8\columnwidth]{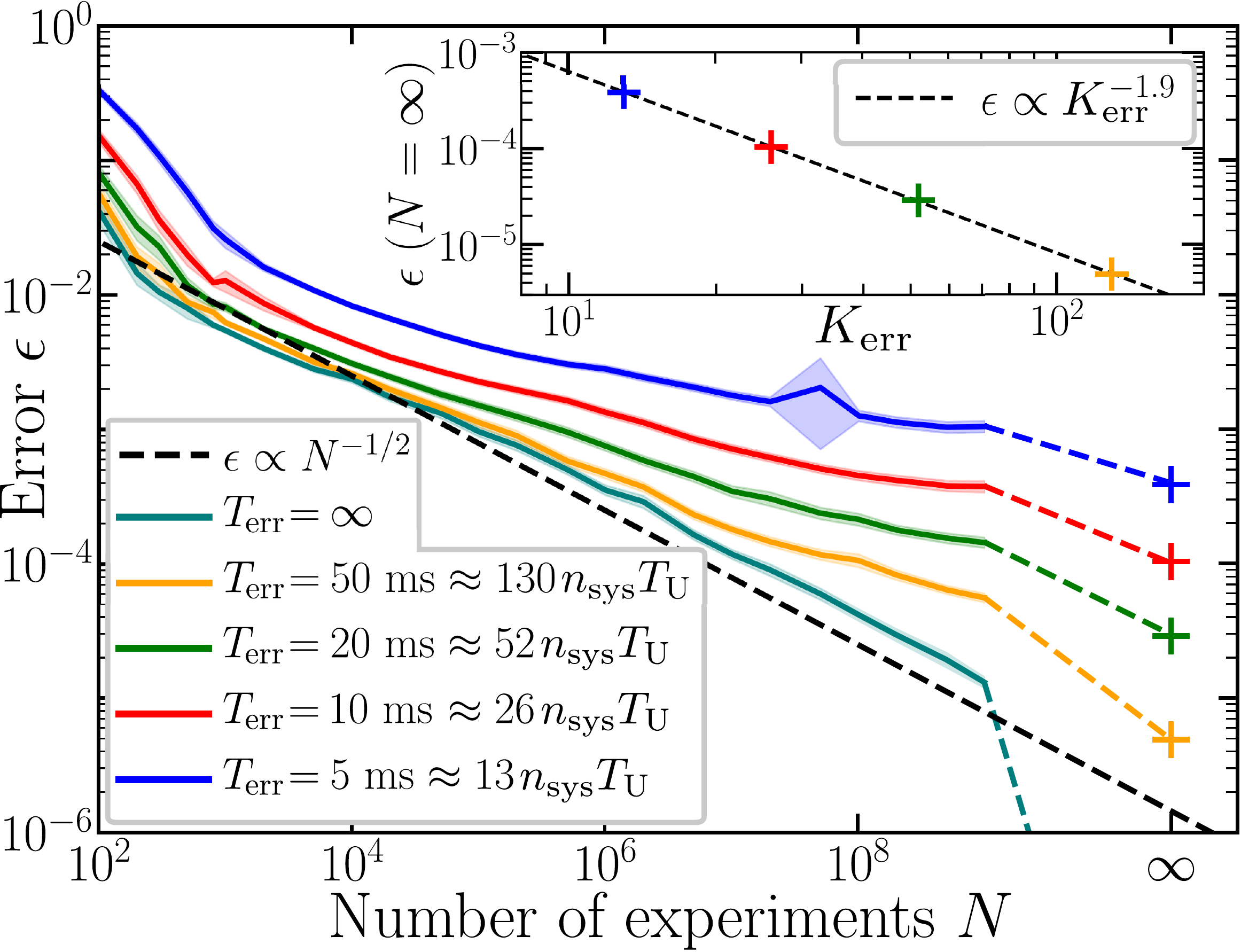}
\caption{\label{fig:TSE_compensation}Performance of time-series estimator with compensation techniques (described in text). Shaded regions denote $95\%$ confidence intervals (averaged over $200$ QPE simulations). Final crosses show the performance in the absence of any sampling noise (teal cross is at approximately $10^{-10}$), i.e. in the limit $N\rightarrow\infty$; dashed lines are present to demonstrate this limit. (inset) Plot of error without sampling noise as a function of the decoherence time $T_{\mathrm{err}}$. Y-axis corresponds to y-axis on main plot (as color-coded).}
\end{figure}

Adjusting the time-series estimator to use only $g(k)$ for positive $k$ gives approximately $1-2$ orders of magnitude improvement.
In Fig.~\ref{fig:TSE_compensation}, we plot the estimator convergence with this method.
We observe that the estimator is no longer unstable, but the $N^{-1/2}$ convergence is never properly regained.
We may study this convergence in greater deal for this estimator, as we may extract $g(k)$ directly from our density-matrix simulations, and thus investigate the estimator performance in the absence of sampling noise (crosses on screen).
We note that similar extrapolations in the absence of noise, or in the presence of depolarizing noise (when compensated) give an error rate of around $10^{-10}$, which we associate to fixed-point error in the solution to the least squares problem (this is also observed in the curve without noise in Fig.~\ref{fig:TSE_compensation}).
Plotting this error as a function of $K_{\rm err}$ shows a power-law decay - $\epsilon\propto K_{\rm err}^{-\alpha} \propto T_{\rm err}^{-\alpha}$ with $\alpha=1.9\approx 2$.
We do not have a good understanding of the source of the obtained power law.

\begin{figure}[htb]
\includegraphics[width=0.8\columnwidth]{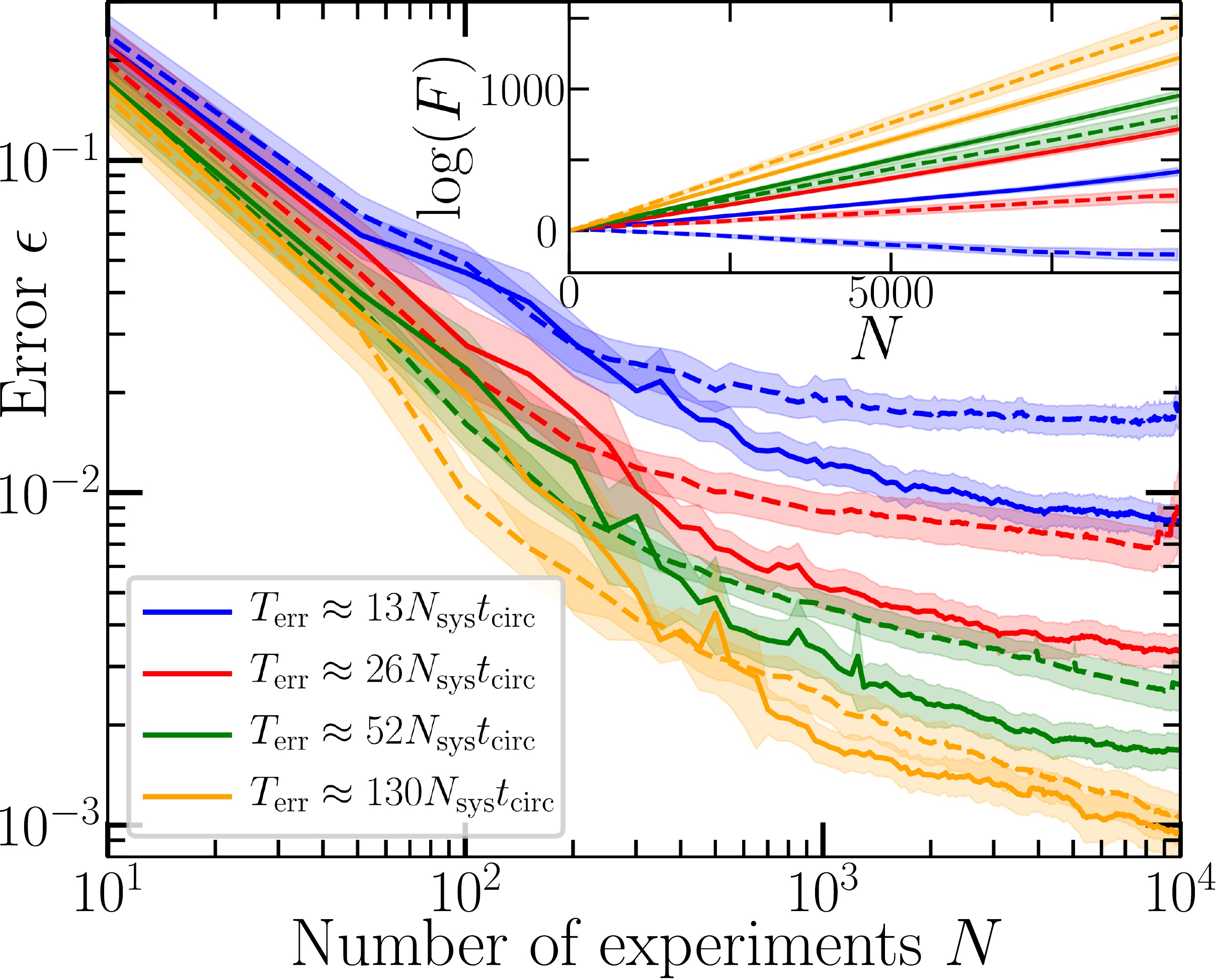}
\caption{\label{fig:singleround_Bayesian_noisy}Performance of single-round Bayesian QPE with four sets of realistic noise using a compensation technique described in the text. Shaded regions are $95\%$ confidence intervals over $200-500$ QPE simulations. (inset) a Bayes factor analysis for the data below. Line color and style matches the legend of the main figure.}
\end{figure}

The same compensation techniques that restored the performance of the Bayesian estimator in the presence of depolarizing noise do not work nearly as well for realistic noise.
Most likely this is due to the fact that the actual noise is not captured by a $k$-dependent depolarizing probability.
In Fig.~\ref{fig:singleround_Bayesian_noisy} we plot the results of using a Bayesian estimator when attempting to compensate for circuit-level noise by approximating it as a depolarizing channel with a decay rate (Eq.~\ref{eq:kdep}) of $K_{\rm err}=T_{\rm err}/T_U\nsys$.
This can be compared with the results of Fig.~\ref{fig:noise_vs_nonoise} where this compensation is not attempted.
We observe a factor $2$ improvement at low $T_{\mathrm{err}}$, however the $N^{-1/2}$ scaling is not regained, and indeed the estimator performance appears to saturate at roughly this point.
Furthermore, at $T_{\mathrm{err}}=50~\mathrm{ms}$, the compensation techniques do not improve the estimator, and indeed appear to make it more unstable.

To investigate this further, in Fig.~\ref{fig:singleround_Bayesian_noisy} (inset) we plot a Bayes Factor analysis of the Bayesian estimators with and without compensation techniques.
The Bayes Factor analysis is obtained by calculating the Bayes Factors
\begin{equation}
F=\prod_{\mathrm{expt}\; n}\frac{P(m_n|M)}{P(m_n|M_0)},
\end{equation}
where $M$ is the chosen Bayesian model (including the prior knowledge), and $M_0$ is a reference model, and $P(m|M)$ is the probability of observing measurement $m$ given model $M$.
As a reference model we take that of random noise - $P(m|M_0)=0.5$.
We observe that at large $T_{\mathrm{err}}$ the Bayes factor with compensation falls below that without, implying that the compensation techniques make the model worse.
We also observe that at very small $T_{\mathrm{err}}$, the estimator makes worse predictions than random noise ($\log(F)<0$).
Despite our best efforts we have been unable to further improve the Bayesian estimator in noisy single-round QPE experiments.

\section{Discussion}

\begin{table}
\begin{tabular}{ c | c | c }\hline & \textbf{Time-series estimator} & \textbf{Bayesian estimator}\\\hline\hline
Speed (scaling) & $O(K)$ & $O(N^2)$ \\\hline
Speed (timing) & \makecell{Processes large datasets\\ in milliseconds} & \makecell{Takes hours to process\\ $10^5$ experiments}\\\hline
Accuracy & \makecell{$\epsilon\propto N^{-1/2}K^{-1}A_0^{-1}\delta^{-2}$\\ demonstrated.} & \makecell{$\epsilon\propto N^{-1/2}K^{-1}$ demonstrated\\ $\epsilon\propto A_0^{-1}\delta^{-2}$ expected.}\\\hline
\makecell{Number of eigenvalues\\ estimated} & \makecell{$100-200$ with\\ relative ease} & Limited to $2-5$\\\hline
\makecell{Improve accuracy\\via classical approximation} & Not obvious & \makecell{Can get speedup via\\ choice of prior\\ (not attempted in this work)}\\\hline
Account for error & Limited ability & Limited ability\\\hline

\end{tabular}
\caption{\label{tab:comp}Table comparing metrics of interest between the two studied estimators. All metrics are implementation-specific, and may be improvable.}
\end{table}

In this work, we have presented and studied the performance of two estimators for quantum phase estimation at low $K$ for different experiment protocols, different systems (in particular those with one vs many eigenvalues), and under simplistic and realistic noise conditions.
These findings are summarized in Table~\ref{tab:comp}.
From our numerical studies, we observe scaling laws for our time-series estimator; we find it first-order sensitive to the overlap $A_0$ between starting state and ground state, second-order sensitive to the gap between the ground state and the nearest eigenstates, and second-order sensitive to the coherence time of the system.
The Bayesian estimator appears to perform comparably to the time-series estimator in all circumstances, and thus should obey similar scaling laws.

We further observe that realistic noise has a worse effect on QPE than a depolarizing channel, for which the effects can largely be mitigated.
We have numerically explored (but not reported) multi-round QPE in the presence of noise.
Since each experiment has multiple outputs, it is harder to adapt the classical data analysis to the presence of noise and our results for realistic noise have not been convincing so far.
Since the performance of multi-round noiseless QPE is already inferior to single-round noiseless QPE, we do not advocate it as a near-term solution, although, for noiseless long circuits it does have the ability to project onto a single eigenstate, which single-round QPE certainly does not.

Despite our slightly pessimistic view of the effect of errors on the performance of QPE, we should note that the obtained error of $10^{-3}$ at $T_{\mathrm{err}}\approx 13 \nsys T_U$ or $K_{\rm err}=13$ would be sufficient to achieve chemical accuracy in a small system.
However, as the energy of a system scales with the number of particles, if we require a Hamiltonian's spectrum to fit in $[-\pi,\pi)$, we will need a higher resolution for QPE, making error rates of $10^{-3}$ potentially too large.
This could potentially be improved by improving the compensation techniques described in the text, applying error mitigation techniques to effectively increase $T_{\mathrm{err}}$, or by using more well-informed prior distributions in the Bayesian estimator to improve accuracy.
All of the above are obvious directions for future work in optimizing QPE for the NISQ era.
Another possible direction is to investigate QPE performance in other error models than the two studied here.
Following Ref.~\cite{KLY:QPE}, we expect SPAM errors to be as innocuous as depolarizing noise.
However, coherent errors can be particularly worrying as they imitate alterations to the unitary $U$.
The time-series estimator is a clear candidate for such a study, due to its ease in processing a large number of experiments and its ability to be studied in the absence of sampling noise.
We also expect that it is possible to combine the time-series estimator with the Heisenberg-limited scaling methods of Refs.~\cite{higg:PE,KLY:QPE} so as to extend these optimal methods to the multiple-eigenvalue scenario with $\neig >1$ eigenvalues, and that these methods could be extended to analog or ancilla-free QPE settings such as described in Ref.~\cite{KLY:QPE}.

In this work we do not compare the performance of quantum phase estimation with purely classical methods. Let's assume that we have a classical efficient representation of the starting state $\Psi$ and one can efficiently calculate ${\rm Tr} \Hh^k |\Psi\>\<\Psi|$ for $k=1, \ldots, K$ with $K=O(1)$ (for fermionic Gaussian starting states and fermionic Hamiltonians this is possible as a single fermionic term in $\Hh^k$ can be estimated as the Pfaffian of some matrix). Then, if there are at most $K=O(1)$ eigenstates in this initial state, the time-series method would allow us to extract these eigenvalues efficiently. Thus in this setting and under these assumptions quantum phase estimation would not offer an exponential computational advantage.

\section*{Acknowledgements}
The authors would like to thank Viacheslav Ostroukh for assistance with quantum simulation, Lucas Visscher for assistance with molecular simulation, Chris Granade for advice on Bayesian techniques, Detlef Hohl and Shell for useful discussions, and Carlo Beenakker, Leonardo DiCarlo, Nathan Wiebe, Ryan Babbush, Jarrod McClean, Yuval Sanders, Xavier Bonet, Sonika Johri and Francesco Buda for advice and feedback on the project. The work by T. E. O'Brien was supported by the Netherlands Organization for Scientific Research (NWO/OCW) and an ERC Synergy grant. The work by B. M. Terhal was supported by ERC grant EQEC No. 682726. The work by B. Tarasinski was supported by the Office of the Director of National Intelligence (ODNI), Intelligence Advanced Research Projects Activity (IARPA), via the U.S. Army Research Office grant W911NF-16-1-0071. The views and conclusions contained herein are those of the authors and should not be interpreted as necessarily representing the official policies or endorsements, either expressed or implied, of the ODNI, IARPA, or the U.S. Government. The U.S. Government is authorized to reproduce and distribute reprints for Governmental purposes notwithstanding any copyright annotation thereon.

\appendix

\section{Derivation of the identity in Eq.~(\ref{eq:chi})}
\label{app:equal}

One first writes for $0 \leq k\leq K/2$:
\begin{align}
\sum_jA_j\exp(i k\phi_j)=\sum_{\mathbf{m}, \mathbf{n}} \Pi_{i=1}^k  [(-1)^{m_i}-i (-1)^{n_i} ] \times \nonumber \\
\mathbb{P}(m_1, \ldots, m_{K/2}, n_1, \ldots n_{K/2}|\boldsymbol{\phi},\mathbf{A})
\label{eq:expon}
\end{align}
where $\mathbb{P}(m_1, \ldots, m_{K/2}, n_1, \ldots n_{K/2}|\boldsymbol{\phi},\mathbf{A})$ is the probability for a specific series of outcomes $m_1,\ldots, m_{K/2}$ for $\beta=0$ and $n_1,\ldots, n_{K/2}$ for $\beta=\pi/2$.
To see that the above is true, note that it is quickly true for $\neig=1$ by using Eq.~(\ref{eq:gk_from_Pfunc}) for $g(1)$.  By linearity on the left and right hand side it then holds generally.

Since the order of the outcomes of the rounds does not matter, i.e. $\mathbb{P}(m_1, \ldots, m_{K/2}, n_1, \ldots n_{K/2}|\boldsymbol{\phi},\mathbf{A})$ only depends on the Hamming weights $\frak{m}=|\mathbf{m}|$ and $\frak{n}=|\mathbf{n}|$, we can symmetrize the coefficient over permutations of the rounds and replace $\mathbb{P}(m_1, \ldots, m_{K/2}, n_1, \ldots n_{K/2}|\boldsymbol{\phi},\mathbf{A})$ by $\mathbb{P}(\frak{m},\frak{n}|\boldsymbol{\phi},\mathbf{A})/({{K/2} \choose \frak{m}} {{K/2} \choose \frak{n}})$. This gives the following expression for $\chi_k(m,n)$:
\begin{align*}
\chi_k(\frak{m},\frak{n}) =&\frac{1}{((K/2)!)^2} \sum_{\pi_1 \in S_{K/2}, \pi_2\in S_{K/2}} \\& \prod_{i=1}^k ((-1)^{m_{\pi_1(i)}}-i (-1)^{n_{\pi_2(i)}}),
\end{align*}
where $m_i$ is the $i$th bit of a bitstring with Hamming weight $\frak{m}$ (and similarly $n_i$), and $S_{K/2}$ is the symmetric group of permutations. We can expand this last expression as 
\begin{align*}
\chi_k(\frak{m},\frak{n})& =& \sum_{k=0}^l {k \choose l} (-i)^{k-l} \rho(l,\frak{m}) \rho(k-l,\frak{n}) \\
\rho(l,\frak{m})& =& \frac{1}{(K/2)!}\sum_{\pi} (-1)^{m_{\pi(1)}}\ldots (-1)^{m_{\pi(l)}} \\
& =& -1+\frac{2}{(K/2)!} \sum_{\pi:m_{\pi(1)}\ldots m_{\pi(l)} \mbox{\tiny is even}}1
\end{align*}
The sum $\sum_{\pi:m_{\pi(1)}\ldots m_{\pi(l)} \mbox{\tiny is even}}$ can be written as a sum over  permutations such that $m_{\pi(1)}\ldots m_{\pi(l)}$ has Hamming weight $2p$ with $p=0,1,\ldots \lfloor l/2 \rfloor$. Then one counts the number of permutations of a $K/2$-bitstring of Hamming weight $\frak{m}$ such that some segment of length $l$ has Hamming weight $2p$ which equals 
${\frak{m} \choose 2p}{K/2-\frak{m} \choose l-2p}\;l!\;(K/2-l)!$.  All together this leads to $\chi_k(\frak{m},\frak{n})$ in Eq.~(\ref{eq:chi}). It is not clear whether one can simplify this equality or verify it directly using other combinatorial identities or (Chebyshev) polynomials.

\section{Variance calculations for time-series estimator}\label{app:calc_variance_full}

For the case of estimating a single eigenvalue using single-round QPE with the time-series estimator, one can directly calculate the error in the estimation.
In this situation, our matrices $G_0$ and $G_1$ are column vectors,
\begin{align}
G_0^T&=(g(-K),g(-K+1),\ldots,g(K-1)),\\ G_1^T&=(g(-K+1),g(-K+2),\ldots,g(K)).
\end{align}
The least-squares solution for $\frak{T}$ is then
\begin{equation}
\frak{T}=(G_0^\dag G_0)^{-1} G_0^\dag G_1=\frac{\sum_{k=-K}^{K-1} g^*(k)g(k+1)}{\sum_{k=-K}^{K-1}g^*(k)g(k)}.
\end{equation}
For a single frequency, $g(k)=e^{ik\phi}$, and immediately $\frak{T}=e^{i\phi}$. However, we estimate the real and imaginary components of $g(k)$ separately. Let us write in terms of our independent components
\begin{equation}
\frak{T}=\frak{T}_r+i\frak{T}_i,\hspace{2cm} g(k)=g^0_k+ig^1_k,
\end{equation}
remembering that $g^0_k=g^0_{-k}$ and $g^1_k=-g^1_{-k}$ (i.e.~the variables are correlated).
Our target angle $\phi=\tan^{-1}\frak{T}_i/\frak{T}_r$, and so we can calculate
\begin{align}
\mathrm{Var}(\phi)&=\sum_{a,k}{\left[\frac{\partial\phi}{\partial g^a_k}\right]}^2\mathrm{Var}[g^a_k]\nonumber\\
&={\left[\frac{1}{\frak{T}_r^2+\frak{T}_i^2}\right]}^2\sum_{a,k}\left[\frak{T}_r\frac{\partial \frak{T}_i}{\partial g_k^a} - \frak{T}_i\frac{\partial \frak{T}_r}{\partial g_k^a}\right]^2\mathrm{Var}[g^a_k].\label{eq:varphidef}
\end{align}
Let us expand out our real and imaginary components of $\frak{T}$:
\begin{align}
\frak{T}_r&=\frac{\sum_{k=-K}^{K-1}(g^0_k g^0_{k+1}+g^1_k g^1_{k+1})}{\sum_{k=-K}^{K-1}(g^0_k)^2+(g^1_k)^2},\\ \frak{T}_i&=\frac{\sum_{k=-K}^{K-1}(g^0_k g^1_{k+1}-g^0_k g^1_{k+1})}{\sum_{k=-K}^{K-1}(g^0_k)^2+(g^1_k)^2}
\end{align}
Then, we can calculate their derivatives as (recalling again that $g^0_k=g^0_{-k}$ and $g^1_k=g^1_{-k}$)
\begin{align}
\frac{\partial \frak{T}_r}{\partial g_k^a}&=\frac{2}{1+\delta_{k,0}}\left[\frac{(1-\delta_{k,K})g_{k+1}^a + g_{k-1}^a - 2\frak{T}_rg_k^a}{\sum_{k=-K}^{k+1}((g_k^0)^2+(g_k^1)^2)}\right]\label{eq:gk1}\\
\frac{\partial \frak{T}_i}{\partial g_k^a}&=\frac{2(-1)^a}{1+\delta_{k,0}}\left[\frac{(1-\delta_{k,K})g_{k+1}^{1-a} - g_{k-1}^{1-a} - 2\frak{T}_ig_k^a}{\sum_{k=-K}^{k+1}((g_k^0)^2+(g_k^1)^2)}\right].\label{eq:gk2}
\end{align}
Substituting in for $g_k^a$, we find that everything precisely cancels when $k\neq K$!
\begin{align}
\frac{\partial \frak{T}_r}{\partial g_k^0}=-\frac{\partial \frak{T}_i}{\partial g_k^1}&=-2\delta_{k,K}\frac{\cos((K+1)\phi)}{\sum_{k=-K}^{k+1}((g_k^0)^2+(g_k^1)^2)}\\
\frac{\partial \frak{T}_i}{\partial g_k^0}=\frac{\partial \frak{T}_r}{\partial g_k^1}&=-2\delta_{k,K}\frac{\sin((K+1)\phi)}{\sum_{k=-K}^{k+1}((g_k^0)^2+(g_k^1)^2)}.
\end{align}
Our variance is then
\begin{align}
&\mathrm{Var}(\phi)=\left[\frac{2}{(\frak{T}_r^2+\frak{T}_i^2)\sum_{k=-K}^{k+1}((g_k^0)^2+(g_k^1)^2)}\right]^2\times\nonumber\\&\left\{\mathrm{Var}[g_K^0]\left(-\cos(\phi)\sin((K+1)\phi)+\sin(\phi)\cos((K+1)\phi)\right)^2\right.\nonumber\\&\left.+\mathrm{Var}[g_K^1]\left(\cos(\phi)\cos((K+1)\phi)+\sin(\phi)\sin((K+1)\phi)\right)^2\right\}\nonumber\\
&=\left[\frac{1}{K}\right]^2\left\{\mathrm{Var}[g_K^0]\sin^2(K\phi)+\mathrm{Var}[g_K^1]\cos^2(K\phi)\right\}.
\end{align}
If $g_K^a$ is estimated with $N$ shots, we expect $\mathrm{Var}[g_K^0]=\frac{1}{N}$, and
\begin{equation}
\mathrm{Var}(\phi)\propto\frac{1}{K^2N}.
\end{equation}

As described in Sec.~\ref{sec:time_mr}, for multi-round experiments we weight the least-squares inversion as per Eq.~(\ref{eq:weight_function}).
This weighting adjusts the $g_k^a$ values in Eqs.~(\ref{eq:gk1},\ref{eq:gk2}) so that $\frac{\partial\phi}{\partial g_k^A}$ is no longer zero when $k<K$.
The sum over $k$ in Eq.~(\ref{eq:varphidef}) then lends an extra factor of $K$ to the variance, reducing it to
\begin{equation}
\mathrm{Var}(\phi)\propto\frac{1}{KN}.
\end{equation}

\section{Fourier representation for Bayesian updating}\label{App:BayesFourier}
For simplicity, we first consider when the starting state is a simple eigenstate $|\phi_j \rangle$.
After each multi-round experiment we would like to update the probability distribution 
$P(\phi_j=\phi)$, i.e. $P_n(\phi)=\frac{P_{\mathbf{k},\boldsymbol{\beta}}(\mathbf{m}|\phi)}{P(\mathbf{m})}P_{n-1}(\phi)$. We will represent the $2\pi$-periodic probability distribution $P_n(\phi)$ by a Fourier series with a small number of Fourier coefficients $\nfreq$ which are updated after each experiment, that is, we write
\begin{equation}
P(\phi)=p_0+\sum_{j=1}^{\nfreq-1}\left(p_{2j-1}\sin(j\phi)+p_{2j}\cos(j\phi)\right) \equiv \boldsymbol{p}.\label{eq:p_def}
\end{equation}
We thus collect the coefficients as a $\nfreq$-component vector $\boldsymbol{p}$. The Fourier representation has the advantage that integration is trivial i.e. $\int_{-\pi}^{\pi} P(\phi)d\phi=2\pi p_0$ so that the probability distribution is easily normalized. In addition, the current estimate $\tilde{\phi}$ is easy to obtain:
\begin{align}
\tilde{\phi}=\mathrm{arg}(\<e^{i\phi}\>_P)=\mathrm{arg}(p_2+ip_1).
\label{eq:estimate-eq}
\end{align}
Another observation is that the Holevo phase variance is easily obtained from this Fourier representation as
\begin{align}
{\rm Var}(P(\phi))=\frac{1}{|\<e^{i\phi}\>_P|^2}-1=\frac{1}{\pi^2(p_2^2+p_1^2)}-1.
\end{align}
Note that this is the Holevo phase variance of the posterior distribution of a single simulation instance.
By comparison, in Fig.~\ref{fig:single_shot} we have calculated the same quantity over repeat simulations.
However, in general we find the two to be equivalent.

The other advantage of the Fourier representation is that a single-round in an experiment is the application of a sparse matrix on $\mathbf{p}$. One has $P(\phi) \rightarrow P_{k_r,\beta_r}(m_r|\phi)P(\phi)=\cos^2(k_r\phi/2+\gamma/2)P(\phi)$, where $\gamma=\beta_r+m_r\pi$ which is equivalent to 
\begin{align}
\mathbf{p} \rightarrow \frac{1}{2}\mathbf{p}+\frac{1}{4}\cos(\gamma)M^0(k_r)\mathbf{p}+\frac{1}{4}\sin(\gamma)M^1(k_r)\mathbf{p}.\label{eq:update_rule1}
\end{align}
The coefficients of the update matrices $M^{0,1}(k_r)$ can be simply calculated using the double angle formulae and employing
\begin{align}
&\cos^2(k\phi/2+\gamma/2)\cos(j\phi)\nonumber\\&=\frac{1}{2}\cos(j\phi)+\frac{1}{4}\cos(\gamma)\left(\cos((j+k)\phi)+\cos((j-k)\phi)\right)\nonumber~\\
&+\frac{1}{4}\sin(\gamma)\left(\sin((j-k)\phi)-\sin((j+k)\phi)\right),
\end{align}
and
\begin{align}
&\cos^2(k\phi/2+\gamma/2)\sin(j\phi)\nonumber\\&=\frac{1}{2}\sin(j\phi)+\frac{1}{4}\cos(\gamma)\left(\sin((j+k)\phi)+\sin((j-k)\phi)\right)\nonumber~\\
&+\frac{1}{4}\sin(\gamma)\left(\cos((j+k)\phi)-\cos((j-k)\phi)\right).
\end{align}
The matrices $M^a(n)$ are then calculated from the above equations. When $j>k$, we have
\begin{align*}
&{[M^0(k)]}_{2j+2k,2j}=1,\;\;\; {[M^0(k)]}_{2j-2k,2j}=1,\\
&{[M^0(k)]}_{2j+2k-1,2j-1}=1,\;\;\; {[M^0(k)]}_{2j-2k-1,2j-1}=1,\\
&{[M^1(k)]}_{2j+2k-1,2j}=-1,\;\;\; {[M^1(k)]}_{2j-2k-1,2j}=1,\\
&{[M^1(k)]}_{2j+2k,2j-1}=1,\;\;\; {[M^1(k)]}_{2j-2k,2j-1}=-1,
\end{align*}
When $j\leq k$, we have to account for the sign change in $\sin((j-k)\phi)$:
\begin{align*}
&{[M^0(k)]}_{j+2k,j}=1,\;\;\; {[M^0(k)]}_{2k-2j,2j}=1,\\ &{[M^0(k)]}_{2k-2j-1,2j-1}=-1\\
&{[M^0(k)]}_{2k,0}=-2,\;\;\; {[M^0(k)]}_{4k-1,2k-1}=1\\
&{[M^1(k)]}_{2j+2k-1,2j}=-1,\;\;\; {[M^1(k)]}_{2k-2j-1,2j}=-1,\\&{[M^1(k)]}_{2j+2k,2j-1}=1,\;\;\; {[M^1(k)]}_{2k-2j,2j-1}=-1,\\
&{[M^1(k)]}_{2k-1,0}=2,\;\;\; {[M^1(k)]}_{4k-1,2k}=1.
\end{align*}

For a multi-round experiment with $R$ rounds, one thus applies such sparse matrices to the vector $\mathbf{p}$ $R$ times.
Note that each round with given $k_r$ requires at most $k_r$ more Fourier components, hence an experiment with at most $K$ controlled-$U$ applications adds at most $K$ Fourier components.
Thus, when the total number of unitary rotations summed over all experiments $K_{\rm tot}=\sum_n\sum_rk_r >\nfreq$, our representation of the distribution is no longer accurate.
When $K_{\mathrm{tot}}\leq \nfreq$ on the other hand, it will be accurate. 

\subsection{Bayesian updating for multi-eigenvalue starting state}
\label{App:BayesFourier-multi}

In this section we detail the method by which we store the distributions $P_n^{j}(\phi_j)$ and $P_n^{\rm red}(\mathbf{A})$ of Eq.~(\ref{eq:Indep_distribution}) and perform the Bayesian update of Eq.~(\ref{eq:Bayesrule}). We do so by representing the marginal probabilities $P_n^j(\phi_j)$ by a Fourier series with a small number of Fourier coefficients which are updated after each experiment as shown in the previous section. We assume that there are most $\neig$ coefficients $A_j >0$ and thus $\neig$ $\phi_j$.
 
From our independence assumption, individual updates of $P^{j}(\phi_j)$ may be calculated by integrating out the other unknown variables in Eq.~(\ref{eq:Bayesrule}):
\begin{equation}
P_{n}^{j}(\phi_j)=\int\left(\prod_{l\neq j} d\phi_l P_{n-1}^{l}(\phi_l)\right) \int d\mathbf{A}\; P_{n-1}^{\rm red}(\mathbf{A}) P_{\mathbf{k},\boldsymbol{\beta}}(\mathbf{m}|\boldsymbol{\phi},\mathbf{A})P_{n-1}^{j}(\phi_j).
\end{equation}
Expanding the conditional probability of Eq.~(\ref{eqn:QPE_probability_noerror}) and rewriting leads to the form
\begin{equation}
P_{n}^{j}(\phi_j)=\frac{1}{P_{\mathbf{k},\boldsymbol{\beta}}(\mathbf{m})}\left(C+B_j\prod_r P_{k_r,\beta_r}(m_r|\phi_j)\right)P_{n-1}^{j}(\phi_j),
\end{equation}
with 
\begin{align*}
C = \sum_{k \neq j} B_k \int d\phi_k P_{n-1}^{k}(\phi_k)\prod_r P_{k_r,\beta_r}(m_r|\phi_k),
\end{align*}
and $B_j =\int d\mathbf{A}\; P_{n-1}^{\rm red}(\mathbf{A}) A_j$. Here we have used that $\int d\phi_l P^l_{n-1}(\phi_l)=1$. One can concisely write $B_j$ as the components of a vector ${\bf B}$.
Computing Eq.~(\ref{eq:Bayesrule}) then involves creating an `update' distribution for each $\phi_j$, calculating the integral of each distribution, and then forming the new distribution from a weighted sum from the `update' distributions.

Calculating the distribution $P_n^{\rm red}(\mathbf{A})$ is complicated slightly by the restriction that $\sum_j A_j =1, A_j \geq 0$, meaning that we cannot assume the distribution of individual $A_j$ terms is uncorrelated. The marginal probability distribution equals
\begin{equation}
P_{n}^{\rm red}(\mathbf{A})=\frac{P_{n-1}^{\rm red}(\mathbf{A})}{P_{\mathbf{k},\boldsymbol{\beta}}(\mathbf{m})}\sum_j A_j \int d\phi_j P_{n-1}^{j}(\phi_j)\prod_r P_{k_r,\beta_r}(m_r|\phi_j).\label{eq:expanded_form_phij}
\end{equation}
or
\begin{equation}
P_n^{\rm red}(\mathbf{A})=\frac{P_{n-1}^{\rm red}(\mathbf{A})}{P_{\mathbf{k},\boldsymbol{\beta}}(\mathbf{m})}\mathbf{A}\cdot\mathbf{q}_{n-1},
\end{equation}
where the $j$th component $(q_{n-1})_j$ is the integral
\begin{equation}
(q_{n-1})_j=\int d\phi_j P_{n-1}^{j}(\phi_j)\prod_r P_{k_r,\beta_r}(m_r|\phi_j).
\end{equation}
As $\mathbf{A}$ only enters our estimation through the vector $\mathbf{B}=(B_0,\ldots,B_{\neig})$, we only need approximate this value. Assuming we know the marginal probabilities $P_{n}(\phi_j)$ for all experiments $n=1,\ldots,N$, we can estimate $\mathbf{B}$ after all experiments by the maximum likelihood value $\mathbf{A}^{(\max)}$,

\begin{align*}
\mathbf{A}^{(\max)}_{N}&=\underset{\mathbf{A}}{{\rm arg max}} f(\mathbf{A})\\f(\mathbf{A})&=\log\left(P_{\rm prior}(\mathbf{A})\prod_{n=1}^{N}\mathbf{A}\cdot\mathbf{q}_n\right)\\
&=\log(P_{\rm prior}(\mathbf{A}))+\sum_{n=1}^{N}\log(\mathbf{A}\cdot\mathbf{q}_n)\label{eq:argmax}.
\end{align*}
Evaluating this equation for up $N=1000$ experiments, taking $\nfreq=10000$ frequency components of $\neig=2$ eigenvalues takes less than a second on a laptop using a method such as sequential least-squares programming~\cite{Kra94}.
However, beyond this it becomes fairly computationally intensive. Thus, after $N > 100$ experiments have been performed, we switch to a local optimization method.
We determine the optimal $\mathbf{B}_n$ after $n$ experiments from its prior value $\mathbf{B}_{n-1}$ via a single step of an approximate Newton's method, that is, we take
\begin{align*}
\mathbf{B}_n =\mathbf{B}_{n-1} -\Pi[\mathbf{H}^{-1}(f(\mathbf{B}_{n-1}))\; (\vec{\nabla} f)(\mathbf{B}_{n-1})].
\end{align*}
where $\vec{\nabla} f(\mathbf{A})$ is the first derivative of $f$ at $\mathbf{A}$ and $\mathbf{H}$ is the Hessian matrix of $f$, i.e. $H_{ij}=\partial_{A_i}\partial_{A_j}f(\mathbf{A})$. Here $\Pi[\mathbf{A}]$ is the projector onto the plane $\sum_{j=0}^{\neig} A_j=1$ so that the update preserves the normalization.
We have 
\begin{align*}
\partial_{A_i} f(\mathbf{A})=\frac{\partial_{A_i} P_{\rm prior}(\mathbf{A})}{P_{\rm prior}(\mathbf{A})}+\sum_{n=1}^{N} \frac{(q_n)_i}{\mathbf{A}\cdot \mathbf{q}_n}
\end{align*}
We approximate the second term for each step as coming from only from the added term, i.e.
\begin{equation}
\vec{\nabla} f(\mathbf{B}_{n-1}) \approx \frac{\mathbf{q}_{n}}{\mathbf{B}_{n-1} \cdot\mathbf{q}_{n}},
\end{equation}
The Hessian equals 
\begin{equation}
H_{ij}(f(\mathbf{A}))= -\sum_{n=1}^{N}\frac{(q_{n})_i (q_{n})_j}{(\mathbf{A} \cdot \mathbf{q}_{n})^2},
\end{equation}
but we approximate this at the $n$th step
\begin{equation}
H_{ij}^{(n)}(f(\mathbf{B}_{n-1})) \approx H_{ij}^{(n-1)}-\frac{(q_{n})_i (q_{n})_j}{(\mathbf{B}_n \cdot \mathbf{q}_{n})^2}.
\end{equation}
This approximation allows $H$ to be updated without summing over each experiment.

With the above implemented, we observe that our estimator can process data from $N=10,000$ experiments to estimate $\neig=2$ eigenvalues with $N=20,000$ Fourier components within approximately two minutes on a laptop. Unfortunately, this method scales as $N^2$, as the number of frequencies required for accurate estimation grows as the total number of unitaries applied.
 
As the mean, variance and integration calculations only require the first few frequencies of the distribution, it may be possible to reduce this cost by finding approximation techniques for higher frequency components.

\section{Convergence of the (noiseless) time-series analysis in case of multiple eigenvalues.}\label{app:multi_evdata}

\begin{figure}
\includegraphics[width=0.6\columnwidth]{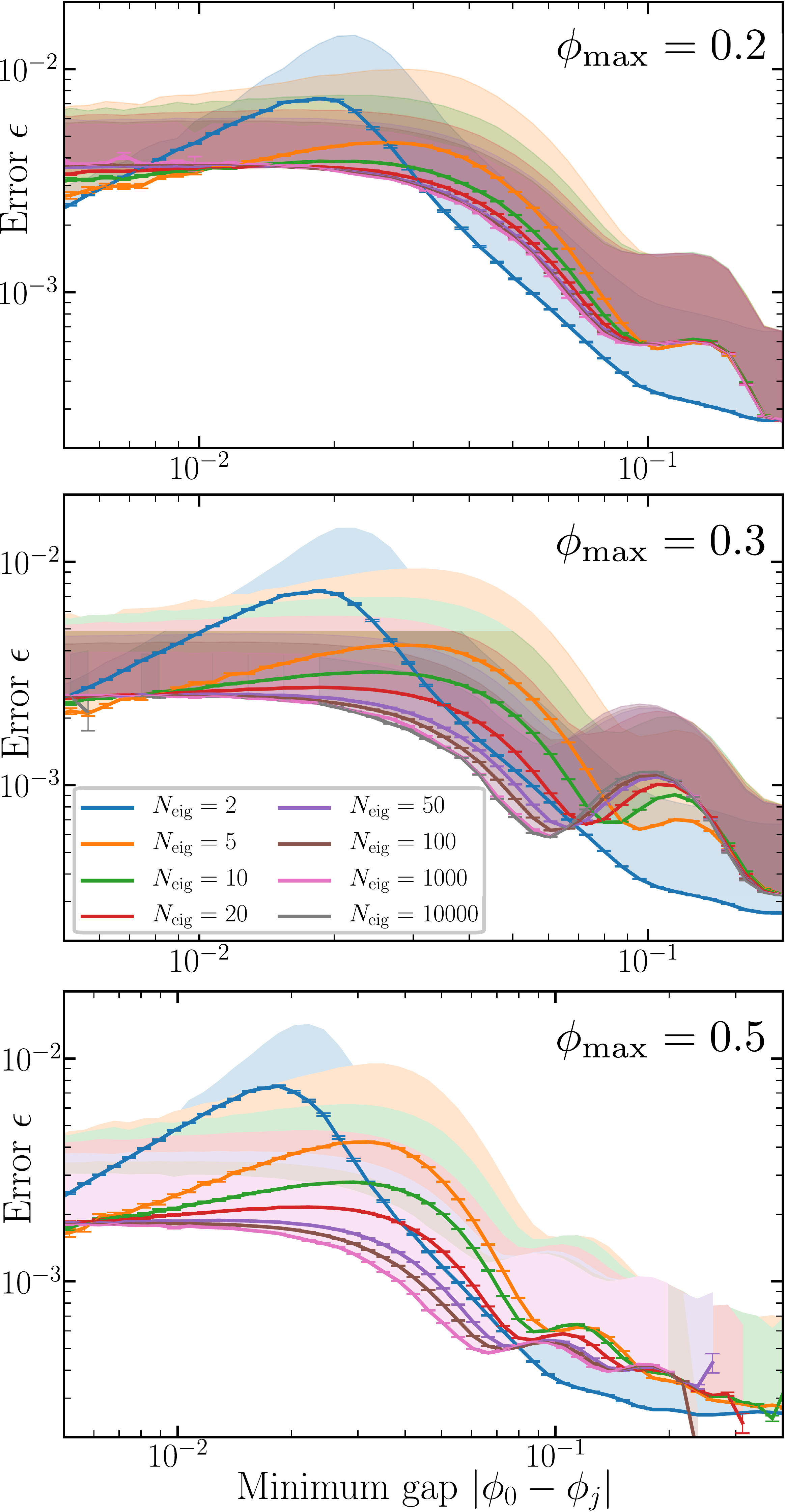}
\caption{\label{fig:changing_phimax_figs}Variations of Fig.~\ref{fig:many_ev}, but with eigenstates $\phi_j$ drawn from a range $[0,\phi_{\max}]$ as labeled. Error bars are $95\%$ confidence intervals for each point, shaded regions denote top $2\sigma$ interval (i.e. region containing the top $2.5\%-50\%$ of the population).}
\end{figure}

In this section we present an expansion of Fig.~\ref{fig:many_ev}, namely Fig.~\ref{fig:changing_phimax_figs}, by drawing the spurious eigenvalues $\phi_j$ from a range closer to the target eigenvalue $\phi_0$.
This negates the drop in estimation error observed in Fig.~\ref{fig:many_ev} that was caused by the majority of eigenvalues lying in region $(c)$ of Fig.~\ref{fig:2ev}.
We observe that for certain gaps $\delta$, multiple eigenvalues confined to a thin region $[\delta,\phi_{\max}]$ can have a worse effect on our ability to estimate $\phi_0$ than that of a single eigenvalue at $\delta$.
However, this loss in accuracy does not get critically worse with the addition of more eigenvalues. Neither is it worse than the worst-possible estimation with two eigenvalues.

\section{Details of realistic simulation}\label{app:sim_details}

In this Appendix we give details of the method for the realistic noisy circuit simulation of Sec.~\ref{sec:realistic_noise}.
Our density-matrix simulator is fairly limited in terms of qubit number, and so we opt to simulate H$_4$ in the STO-3G basis.
This molecule has $8$ spin orbitals and thus requires $9$ qubits for the QPE simulation (with the additional qubit being the ancilla).
We choose $10$ rectangular molecular geometries for the H$_4$ system, parametrized by a horizontal distance $d_x$ and a vertical distance $d_y$ (i.e. the four H atoms are in the positions $(\pm d_x/2,\pm d_y/2, 0)$).
We calculate the Hartree-Fock and full-CI solutions to the ground state using the psi4 package~\cite{Par17} with the openfermion interface~\cite{Mcc17}.
This allows to calculate the true ground state energy $E_0$ for each geometry, and the overlap $A_0$ between the ground state and the Hartree-Fock state, which we choose as our starting state $|\Psi\>$.
Due to symmetry and particle number conservation, $|\Psi\>$ has non-zero overlap with only $8$ eigenstates of the full-CI solution, separated from the ground state by a minimum gap $\delta$.
(When $d_x=d_y$, the true ground state of H$_4$ is actually orthogonal to the Hartree-Fock state, and so we do not include any such geometries in our calculation.)
The full error in our calculation of the energy (at a fixed geometry) is then a combination of three separate contributions: basis set error (i.e.~from the choice of orbitals), Trotter error, and the estimator error studied in this work (which includes error from experimental noise).
The Trotter error $\epsilon_{\mathrm{Trotter}}$ is reasonably large due to our use of only the first-order Suzuki-Trotter approximation $U=\prod_i e^{-i H_i t}\approx e^{-i\Hh t}$.
Higher-order Suzuki-Trotter expansions require longer quantum circuits, which in turn increase the estimator error from experimental noise.
Balancing these two competing sources of error is key to obtaining accurate calculations and a clear target for future study.
In Tab.~\ref{tab:H4params}, we list some parameters of interest for each studied geometry.
We normalize the gap and the Trotter error by the Frobenius norm $\|\Hh\|_F = \sqrt{\mathrm{Trace}[\Hh^{\dag}\Hh]/2^{\nsys}}$, as we chose an evolution time $t=1/\|\Hh\|_F$, making this the relevant scale for comparison with scaling laws and errors calculated in the text.

\begin{table}[htb]
\begin{tabular}{|c|c|c|c|c|c|}\hline
$d_x$ [\r{A}] & $d_y$ [\r{A}] & $E_0$ & $A_0$ & $\delta$/$\|\Hh \|_F$ & $\epsilon_{\mathrm{Trotter}}/\|\Hh \|_F$ \\\hline
0.4 & 0.5 & -0.26 & 0.98 & 0.09 & $3.7\times10^{-4}$ \\\hline
0.6 & 0.7 & -1.46 & 0.94 & 0.17 & $3.1\times10^{-3}$ \\\hline
0.8 & 0.9 & -1.84 & 0.88 & 0.24 & 0.016 \\\hline
1.0 & 1.1 & -1.96 & 0.80 & 0.23 & 0.017 \\\hline
1.2 & 1.3 & -1.98 & 0.71 & 0.18 & 0.013 \\\hline
1.6 & 1.7 & -1.94 & 0.55 & 0.09 & $6.0\times 10^{-3}$ \\\hline
0.2 & 1.8 & 0.32 & 0.996 & 0.67 & $2.0\times 10^{-4}$ \\\hline
0.4 & 1.6 & -1.80 & 0.993 & 1.14 & $2.6\times 10^{-3}$ \\\hline
0.6 & 1.4 & -2.15 & 0.98 & 1.27 & 0.014 \\\hline
0.8 & 1.2 & -2.09 & 0.96 & 0.73 & 0.021 \\\hline
\end{tabular}
\caption{\label{tab:H4params}Parameters of the H$_4$ geometries used in the text. Terms are described in \ref{app:sim_details}. $||\Hh||_F=\sqrt{\mathrm{Trace}[\Hh^{\dagger} \Hh]/2^{\nsys}}$.}
\end{table}

\subsection{Error model and error parameters}\label{App:quantumsim}

\begin{table}[htb]
\begin{tabular}{| l | l | l|l | }
    \hline
    Parameter & Symbol & Standard Value & Scaling \\
    \hline
    Qubit relaxation time & $T_1$ & $30~\mu\mathrm{s}$ & $\lambda$\\
    Qubit dephasing time & $T_2$ & $30~\mu\mathrm{s}$ & $\lambda$\\
    Single-qubit gate time & $T_{\mathrm{sq}}$ & $20~\mathrm{ns}$ & $1$\\
    Two-qubit gate time & $T_{\mathrm{2q}}$ & $40~\mathrm{ns}$ & $1$\\
    In-axis rotation error & $p_{\mathrm{axis}}$ & $10^{-4}$ & $\lambda^{-1}$\\
    In-plane rotation error & $p_{\mathrm{plane}}$  & $5\times 10^{-4}$ & $\lambda^{-1}$\\
    Incoherent flux noise & $A$ & $(1\mu\Phi_0)^2$ & $\lambda^{-1}$\\
    Measurement time & $T_{\mathrm{meas}}$ & $300~\mathrm{ns}$ & $1$\\
    Depletion time & $T_{\mathrm{dep}}$ & $300~\mathrm{ns}$ & $1$\\
    Readout infidelity & $\epsilon_{\mathrm{RO}}$ & $5\times 10^{-3}$ & $\lambda^{-1}$\\
    Measurement induced decay & $p_{\mathrm{d,i}},p_{\mathrm{d,f}}$ & $0.005$, $0.0015$ & $\lambda^{-1}$\\
    \hline
\end{tabular}
\caption{\label{table:parameters} Standard parameters of error models used in density matrix simulation. Table adapted from Ref.~\cite{Obr17} with all parameters taken from therein (with the exception of the $1/f$ flux noise, which is made incoherent as described in text).}
\end{table}

Throughout this work we simulate circuits using an error model of superconducting qubits first introduced in Ref.~\cite{Obr17}.
This captures a range of different error channels with parameters either observed in experimental data or estimated via theory calculations.
All error channels used are listed in Tab.~\ref{table:parameters}, and we will now describe them in further detail.

Transmon qubits are dominated primarily by decoherence, which is captured via $T_1$ and $T_2$ channels~\cite{Nie00}.
Typical $T_1$ and $T_2$ times in state-of-the-art devices are approximately $10-100~\mu\mathrm{s}$.
As other error parameters are derived from experimental results on a device with $T_1=T_2\approx30~\mu\mathrm{s}$, we take these as a base set of parameters~\cite{Bul16,Rol17}.
Single-qubit gates in transmon qubits incur slight additional dephasing due to inaccuracies or fluctuations in microwave pulses.
We assume such dephasing is Markovian, in which case it corresponds to a shrinking of the Bloch sphere along the axis of rotation by a value $1-p_{\mathrm{axis}}$, and into the perpendicular plane by a value $1-p_{\mathrm{plane}}$.
We take typical values for these parameters as $p_{\mathrm{axis}}=10^{-4}$, $p_{\mathrm{plane}}=5\cdot 10^{-4}$~\cite{Obr17}.

Two-qubit gates in transmon qubits incur dephasing due to $1/f$ flux noise.
Assuming that the phase in an ideal C-Phase gate $G={\rm diag}(1,1,1,e^{i \phi}))$ is controlled by adjusting the time of application, this suggests a model for the applied gate which is 
\begin{equation}
G(\delta_{\mathrm{flux}})=\left(\begin{array}{cccc}1&0&0&0\\0&1&0&0\\0&0&e^{i\delta_{\mathrm{flux}}\phi}&0\\0&0&0&e^{i(1+\delta_{\mathrm{flux}}/2)\phi}\end{array}\right),
\end{equation}
where $\delta_{\mathrm{flux}}$ is drawn from a normal distribution around $0$ with standard deviation $\sigma_{\mathrm{flux}}$. One can estimate $\sigma_{\mathrm{flux}}\approx0.01~\mathrm{rad}$ for a typical gate length of $40~\mathrm{ns}$~\cite{Obr17}.
The noise is in general non-Markovian, as $\delta_{\mathrm{flux}}$ fluctuates on longer timescale than a single gate. However, to make the simulation tractable, we approximate it as Markovian. The Pauli transfer matrix of this averaged channel~\cite{Cho12} reads 
\begin{equation}
\Lambda[G]=\int d\delta_{\mathrm{flux}} P(\delta_{\mathrm{flux}})\Lambda[G(\delta_{\mathrm{flux}})],
\end{equation}
where the Pauli transfer matrix of a channel $G$ is given by $\Lambda[G]_{i,j}=\mathrm{Tr}[\sigma_iG\sigma_j]$.

During qubit readout, we assume that the qubit is completely dephased and projected into the computational basis.
We then allow for a $T_{\mathrm{meas}}=300~\mathrm{ns}$ period of excitation and de-excitation (including that from $T_1$-decay), during which the qubit state is copied onto a classical bit. 
This copying is also assumed to be imperfect, with a probability $\epsilon_{\mathrm{RO}}$ of returning the wrong result.
The qubit then has an additional $T_{\mathrm{dep}}=300~\mathrm{ns}$ waiting period before it may participate in gates again (to allow resonator depletion~\cite{Bul16}), over which additional excitation and de-excitation may occur.
Though simple, this description is an accurate model of experimental results.
Typically experiments do not observe measurement-induced excitation to the $|1\>$ state, but do observe measurement-induced decay~\cite{Obr17}.
Typical values of such decay are $0.005$ prior to the copy procedure, and $0.015$ after.

Though reasonably accurate, this error model does fail to capture some details of real experimental systems.
In particular, we do not include leakage to the $|2\>$ state, which is a dominant source of two-qubit gate error. Furthermore, we have not included cross-talk between qubits.

To study the effect of changing noise levels while staying as true as possible to our physically-motivated model, we scale our noise parameters by a dimensionless parameter $\lambda$ such that the contribution from each error channel to the simulation remains constant.
In Tab.~\ref{table:parameters} we show the power of $\lambda$ that each error term is multiplied by during this scaling.
We report $T_{\mathrm{err}}:=T_1=T_2$ in the main text instead of $\lambda$ to make connection to parameters regularly reported in experimental works.

\bibliographystyle{hunsrt}
\bibliography{paper13-NJP.bbl}

\end{document}